\documentclass[pre,floatfix,superscriptaddress,nofootinbib, longbibliography,twocolumn,final]{revtex4-1}

\usepackage{graphicx}
\usepackage[utf8]{inputenc}
\usepackage[english]{babel}
\usepackage{microtype}

\usepackage[boxruled]{algorithm2e}
\SetAlFnt{\footnotesize}
\SetAlCapFnt{\small}
\SetAlCapNameFnt{\small}

\usepackage{amsmath}
\usepackage{amsfonts}
\usepackage{amsthm}
\usepackage{amssymb}
\usepackage{bm} 
\newtheorem{proposition}{Proposition}
\DeclareMathOperator{\E}{\mathop{\mathbb{E}}}

\usepackage{tabularx}
\usepackage{booktabs} 
\usepackage{array} 
\usepackage{dsfont}
\usepackage[normalem]{ulem}

\usepackage[colorlinks, breaklinks,hyperfootnotes = false,citecolor = red]{hyperref} 
\usepackage[usenames]{color}

\usepackage[capitalise]{cleveref} 

\newcommand{\estOmega}[1]{\ensuremath{\widehat{\bm{\Omega}}^{(#1)}}}
\newcommand{\estOmegaNoSup}[0]{\ensuremath{\widehat{\bm{\Omega}}} }
\newcommand{\F}[0]{\ensuremath{\textrm{F}} }

\begin{document}
\title{Hierarchical community structure in networks}

\author{Michael T. Schaub}
\email{michael.schaub@rwth-aachen.de}
\affiliation{Department of Computer Science, RWTH Aachen University}
\author{Jiaze Li}
\affiliation{Department of Data Analytics and Digitalisation, School of Business and Economics, Maastricht University}
\author{Leto Peel}
\email{l.peel@maastrichtuniversity.nl}
\affiliation{Department of Data Analytics and Digitalisation, School of Business and Economics, Maastricht University}

\begin{abstract}
	Modular and hierarchical community structures are pervasive in real-world complex systems. 
	A great deal of effort has gone into trying to detect and study these structures.
	Important theoretical advances in the detection of modular have included identifying fundamental limits of detectability by formally defining community structure using probabilistic generative models. 
	Detecting hierarchical community structure introduces additional challenges alongside those inherited from community detection. 
	Here we present a theoretical study on hierarchical community structure in networks, which has thus far not received the same rigorous attention.
	We address the following questions:
    1)~How should we define a hierarchy of communities? 2)~How do we determine if there is sufficient evidence of a hierarchical structure in a network? and 3)~How can we detect hierarchical structure efficiently?
We approach these questions by introducing a definition of hierarchy based on the concept of stochastic externally equitable partitions and their relation to probabilistic models, such as the popular stochastic block model. 
	We enumerate the challenges involved in detecting hierarchies and, by studying the spectral properties of hierarchical structure, present an efficient and principled method for detecting them. 
\end{abstract}

\maketitle

\section{Introduction}
Hierarchical organization has been a central theme in the study of complex systems, dating back to the seminal work of Herbert Simon~\cite{Simon1962}, who observed that a large proportion of complex systems exhibit hierarchical structure. 
Decomposing a complex system into such a hierarchy provides an interpretable summary, or coarse-grained description of the system at multiple resolutions. 
As networks have become ubiquitous for modeling complex systems, these ideas have re-emerged as the identification of hierarchical groups, or \textit{communities}, of nodes within a network~\cite{Clauset2008,blundell2013bayesian, peixoto2014hierarchical,Lyzinski2017}. 
Community detection in networks has received a lot of attention because it can reveal important insights about social~\cite{adamic2005political, cortes2001communities, shai2017case} and biological~\cite{haggerty2014pluralistic, holme2003subnetwork, guimera2005functional, shai2017case} systems, among others. 
A hierarchical description of communities provides the additional utility that it enables a consistent multiscale description, linking the organizational structure of a system across multiple resolutions. 
Hierarchical communities thereby circumvent a prominent issue of community detection, namely, deciding an appropriate resolution~\cite{Reichardt2004,Reichardt2006,Traag2011} or number of communities to detect~\cite{Karrer2011, newman2016equivalence}. 
On the other hand, detecting hierarchical communities inherits, and even exacerbates, many of the theoretical and computational challenges of detecting network communities at a single scale. 
Specifically, major challenges for detecting hierarchical communities are: (i)~how should we define a hierarchy of communities? (ii)~how should we determine if a hierarchical structure exists in a network? and (iii)~how can we detect hierarchical structure efficiently?
Recently, we have seen important developments in the theory of community detection and its limitations~\cite{mossel2016,Mossel2018,abbe2016exact, decelle2011asymptotic,Peele1602548} (see also~\cite{abbe2018community,moore2017computer} for reviews). 
Here we lay the foundations for developing such theory for detecting hierarchical community structure in networks. 

The notion of hierarchy in networks is widespread and has been discussed from a plethora of different perspectives~\cite{Corominas-Murtra2013}. 
For instance, if edges denote some type of flow (e.g., information, data, mass, nutrients, money) this may induce a hierarchy among the nodes~\cite{Clauset2015,Shrestha2018} in which nodes higher up in the hierarchy have more links directed towards nodes at lower levels of the hierarchy (or vice versa, depending on the convention of the directionality).
To be clear, these types of nodal rankings are not the hierarchies we are looking for. 
Rather, we are interested in the hierarchical organization of community structure, i.e.,~communities that are again composed of communities~etc. 
Existing models and methods for detecting hierarchical structure are often constrained to find dense assortative community structures~\cite{ravasz2003hierarchical,Rosvall2011, Lancichinetti2011}. 
Here we consider general probabilistic descriptions of mesoscopic hierarchical group structures, which can be combinations of assortative and disassortive structure.

\begin{figure}
 \includegraphics[width=\columnwidth]{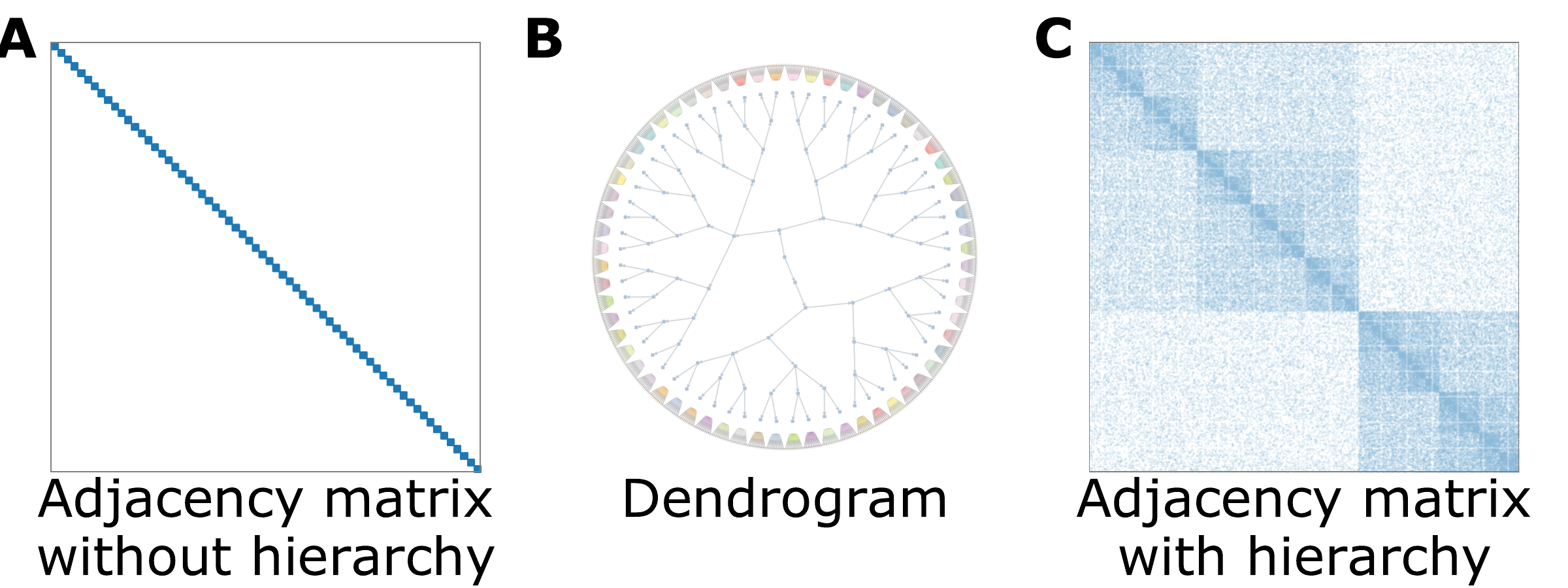}
 \caption{\textbf{A hierarchical model does not guarantee hierarchical community structure.} (A)~An adjacency matrix of a network with 64 groups of fully connected nodes (cliques), each containing ten nodes each.  This network contains an unambiguously ``flat'' partition that contains no hierarchy. (B)~The dendrogram representing the hierarchy  found by detecting communities using a hierarchical model. (C)~An adjacency matrix containing hierarchically structured block densities that is consistent with the dendrogram structure in~(B).}
 \label{fig_modelvshier}
\end{figure}

There are currently many methods available that perform ``hierarchical'' community detection.
Some methods are algorithmically hierarchical~\cite{blondel2008fast, white2005spectral, newman2006modularity} and produce a hierarchy as a by-product and without guarantees of hierarchical structure in the network. 
Yet another class of methods involve fitting a hierarchical model~\cite{peixoto2014hierarchical,blundell2013bayesian,Clauset2008,Lyzinski2017,leskovec2010kronecker}. 
However, in some cases, the design of these models have been motivated by objectives other than detecting hierarchies such as to produce networks with certain statistical properties~\cite{leskovec2010kronecker} or to identify communities beyond the resolution limit~\cite{peixoto2014hierarchical}. 
Consequently, whenever we use one of these approaches (either a hierarchical algorithm or hierarchical model), we run the risk of identifying a hierarchy that has greater complexity than the data can support. 

To demonstrate this point, \Cref{fig_modelvshier} illustrates a network containing 64 cliques that each contain ten nodes. 
It is relatively uncontroversial to suggest that the desired output of a community detection algorithm for this network would be to recover those sixty-four cliques as communities.
Furthermore, because the cliques are structurally identical, any hierarchical grouping is compatible~---~any clique can be swapped with any other, all putative hierarchical configurations are effectively equivalent and there exists no preferred hierarchical grouping.
Na\"ively applying a hierarchical community detection method may produce a hierarchical clustering, as shown in \cref{fig_modelvshier}B. We can consider this detection of superfluous hierarchical levels as analogous to identifying spurious communities in an Erd\H{o}s-R\'enyi random graph.

These issues typically arise when we simply optimize an objective function, e.g., maximizing modularity, likelihood or posterior probability. 
For instance, the maximum a-posteriori solution may contain multiple hierarchical levels that provide a more compact description within the chosen model class, i.e., a ``simpler'' description of the data, and is therefore optimal with respect to chosen objective. 
However, this notion of simplicity conflicts with the intuition that there is no further structure beyond partitioning the network into sixty-four groups. Note that this is not to say that the maximum a-posteriori solution is bad, as it does present a plausible model that is compatible with the observed data, but rather that it presents an unintuitive interpretation of the hierarchical structure in the data.
Some solutions to this problem exist in the realm of Bayesian inference, where we can take an average or form a consensus according to a distribution over solutions.
Such solutions have been successfully demonstrated for both the regular~\cite{zhang2014scalable,riolo2020consistency} and hierarchical~\cite{Clauset2008,peixoto2020revealing} variants of the community detection problem. 
However, these methods of statistical inference can be computationally demanding. 
Previous approaches either employ Markov chain Monte Carlo methods~\cite{Clauset2008,peixoto2014hierarchical}, for which convergence can be slow and difficult to diagnose, or rely on approximate heuristics that scale quadratically with the number of nodes in the network~\cite{blundell2013bayesian} and are thus limited to relatively small networks.  
Recently, however, fast spectral methods based on the non-backtracking~\cite{Krzakala2013} and Bethe Hessian~\cite{Saade2014} operators have been developed that can efficiently detect communities right down to the theoretical limit of detectability~\cite{Krzakala2013}. 

Spectral algorithms have also been studied in the context of hierarchical communities.
\cite{Lyzinski2017,li2020hierarchical,lei2020consistency,balakrishnan2011noise}.
For instance, White and Smyth~\cite{white2005spectral} and Newman~\cite{newman2006modularity} present spectral algorithms based on the modularity matrix that recursively bipartition a network. These algorithms output a hierarchy in the form of a binary dendrogram, but with the goal of simply recovering a single partition of the network. 
Lyzinski~\cite{Lyzinski2017} analyse the performance of spectral algorithms under a hierarchical generative model based on a random dot product graph model.
Local spectral algorithms have also been shown to provide good solutions when optimizing conductance based scores~\cite{mahoney2012local,jeub2015think,kloster2014heat}, which are of particular interest for very large graphs in case we do not need to partition the graph as a whole.

In this work, we propose a number of important theoretical advances for the detection of hierarchical communities. 
We first provide a definition of hierarchical communities by introducing the concept of stochastic externally equitable partitions and drawing a connection to the popular stochastic block model and various node equivalence classes (Section~\ref{sec:hier_def}). 
Second, we discuss specific challenges that pertain to the detection of hierarchical communities with a specific focus on identifiability issues, which demonstrate that even well-defined hierarchies do not have a unique representation (Section~\ref{sec:ambiguous_hier}). 
Then we turn our attention to the spectral properties of networks with planted hierarchical structures.  
Using these spectral properties, we develop an efficient method for detecting if a hierarchy of communities exists and identifying a hierarchy when it is present (Section~\ref{sec:hier_via_spectral_methods}).
We conduct numerical experiments that demonstrate the efficacy of our approach on synthetic networks (Section~\ref{sec:num_experiments}) and real-world networks (Section~\ref{sec:real_experiments}). Finally we conclude with a discussion of possible extensions of our work, including theoretical consideration and extensions to other type of network models.

\section{Hierarchical structure in networks}
\label{sec:hier_def}
\begin{figure}
 \includegraphics[width=\columnwidth]{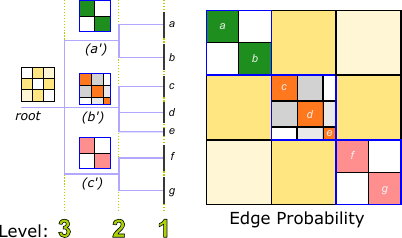}
 \caption{\textbf{Schematic representation of the link probabilities in a hierarchical graph model}. 
 The dendrogram on the left represents the hierarchical partition associated with the hierarchical organization of the edge probabilities on the right.
 At each branch of the dendrogram the link probabilities pattern within the diagonal blocks is refined, as indicated by the smaller block patterns on branch (a')-(c').
 The resulting in an overall edge probability pattern is shown on the right.}
 \label{fig_hiergen}
\end{figure}

Before we can detect hierarchies, it is necessary to define precisely what we mean by a hierarchical structure. 
Any hierarchy can be represented as a rooted tree, sometimes referred to as a dendrogram. 
The root of this tree represents the group of all nodes in the network.
    Starting from the root, at each branch of the dendrogram each parent group is partitioned into child subgroups (see~\Cref{fig_hiergen} for a schematic example).

In hierarchical community detection, as considered here, we aim to identify groups of similar nodes in a network, such that with each further subdivision of the nodes, the resulting groups should contain increasingly similar nodes.
Each subgroup should therefore also have inherited certain similarities from its parent group. 
A relevant way to define similarity is in terms of \textit{stochastically equivalent} nodes, i.e., groups of nodes $r$ and $s$ such that any node in group $r$ has the same probability, $\bm{\Omega}_{rs}$, of linking to any node in group $s$.  
In this setting one can represent the community structure of a network with $n$ nodes using the stochastic block model~(SBM)~\cite{holland1983stochastic, nowicki2001estimation}. 
The SBM defines the probability of a link existing between two nodes depending on their community assignment. We represent this group assignment as a group indicator matrix $\bm{H}\in\{0,1\}^{n\times k}$, in which $\bm H_{ir} = 1$ if node $i$ is assigned to group $r$ and $\bm H_{ir} = 0$ otherwise. Then the probability of nodes $i$ and $j$ being linked is given by,
\begin{equation}
  P[\bm{A}_{ij} = 1] = \bm{H}_{i\bm\cdot}\bm{\Omega}\bm{H}_{j\bm\cdot}^\top \enspace ,
\end{equation}
where $\bm{H}_{i\bm\cdot}$ is the $i$\textsuperscript{th}  row of $\bm{H}$ and $\bm{A}\in \{0,1\}^{n\times n}$ is the adjacency matrix in which $\bm{A}_{ij} = 1$ if there is a link between $i$ and $j$ and $\bm{A}_{ij} = 0$ otherwise. 
Ordering the rows and columns of the adjacency matrix according to the group assignment of nodes allows us to represent $\bm{A}$ as a set of blocks with link densities given by the affinity matrix $\bm{\Omega}$.
Note that in the above description we have allowed self-loops and we will only consider undirected graphs, for simplicity.
Some comments on extensions to directed graphs are provided in the discussion section.

\begin{figure*}
    \centering
    \includegraphics[width=\linewidth]{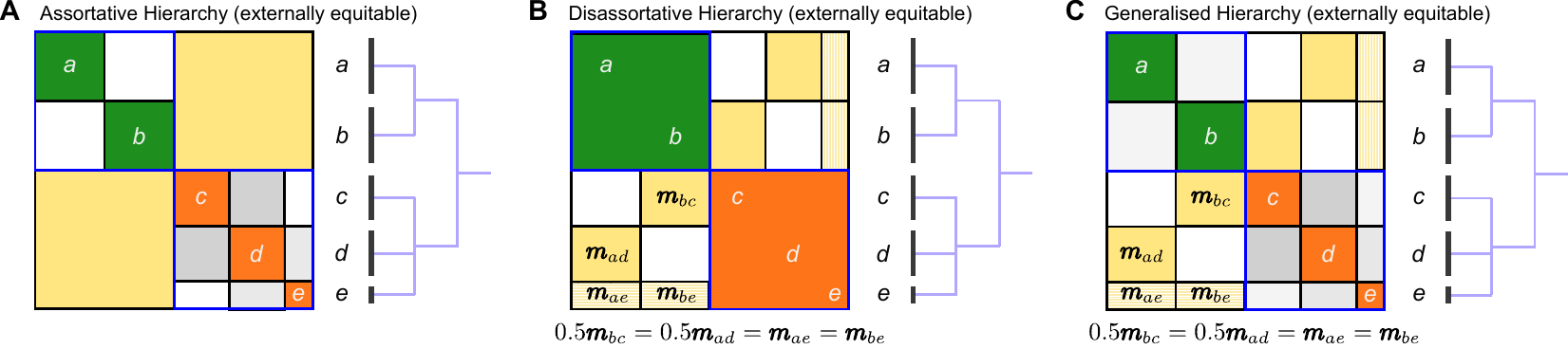}
    \caption{\textbf{The block structure of configurations of hierarchical communities.} (A)~A simple assortative hierarchy of communities in which the refinement of the block structure between levels of the hierarchy occurs along the block diagonal and the off-diagonal blocks have homogeneous density. This type of hierarchy is the most frequently considered in the literature. Although we refer to this structure as an assortative hierarchy, the communities may be disassortative if the off-diagonal blocks are higher density than the diagonal blocks. (B)~A disassortative hierarchy of communities in which the refinement of the block structure between levels of the hierarchy occurs in the off-diagonal blocks. Note that for the disassortative hierarchy to be an externally equitable partition, it must satisfy the stricter constraint that the sum of densities of the refined off-diagonal blocks should be equal along the rows and along the columns, i.e.,
        $\bm m_{ad} +\bm  m_{ae} =
        \bm m_{bc} +\bm m_{be}$ and $\bm m_{ad} = \bm m_{bc} =\bm  m_{ae} +\bm  m_{be}$.  (C)~A generalized hierarchy in which the refinement occurs in both the diagonal and off-diagonal blocks.
}%
    \label{fig:schematic_hier}
\end{figure*}

Based on such an SBM, one way to generate a hierarchy of communities is by recursively subdividing a block into more blocks and describe it as a type of a hierarchical random graph (HRG) model~\cite{Clauset2008} (or its generalized variant~\cite{Peel2015}). 
Figure~\ref{fig_hiergen} illustrates such a hierarchy of communities. 
Starting at the root of the dendrogram in the left of the figure, we generate the total expected number of edges in the network $m$, the number of groups $k^{(1)}$, and a $k^{(1)} \times k^{(1)}$ expected edge count matrix $\bm{m}_{rs}^{(1)}$ that describes how the $m$ links are distributed between the groups, i.e., $\sum_{rs} \bm{m}_{rs}^{(1)} = m$. Note that by convention $\bm{m}_{rr}^{(1)}$ is equal to twice the number of (undirected) edges in group $r^{(1)}$.
The process continues by subdividing each of the groups in the same manner. 
For instance, we can subdivide the $n_r^{(1)}$ nodes in group $r^{(1)}$ by defining an edge count matrix $\bm{m}_{rs}^{(2)}$ that describes how the $\bm{m}_{rr}^{(1)}$ edges are distributed among the subgroups, i.e., $\sum_{rs} \bm{m}_{rs}^{(2)} = \bm{m}_{rr}^{(1)}$. 

Multiple branches may occur simultaneously at the same \textit{level} of the hierarchy, e.g., branches (a'), (b') and (c') occur at the same level in the example in Fig.~\ref{fig_hiergen}. 
We can represent each level $u$ by an assignment of nodes to groups and an affinity matrix, 
$$\bm{\Omega}^{(u)}_{rs} = \frac{\bm m_{rs}^{(u)}}{n_r^{(u)}n_s^{(u)}} \enspace ,$$
of connection probabilities that includes all groups in the network at level $u$. 
In other words, each level may be seen as an SBM that captures a particular resolution of the system.  
Each of the subgroups shares the stochastic equivalence inherited from the parent group, such that all child subgroups of the same parent share the same set of external connection probabilities to other groups. 
Specifically, the probability of a link between two nodes will be governed by the nearest common ancestor in the dendrogram. 

Describing hierarchical communities in this way suggests that we should observe a particular pattern of edge densities in the adjacency matrix when the rows and columns are ordered appropriately. 
We observe such an example in \cref{fig_hiergen}, in which there is a hierarchical refinement of the block structure in the block diagonal of the adjacency matrix and a homogeneous density of edges in the off-diagonal. 
This notion of hierarchical group structure is one of the most common conceptualizations of hierarchical structure encountered in the literature~\cite{Clauset2008, blundell2013bayesian, Lyzinski2017}. We refer to this type of hierarchy as an \emph{assortative hierarchy}.

These assortative hierarchical communities, however, may be limited in their representation of network connection patterns.
For instance, Figure~\ref{fig:schematic_hier}A illustrates an assortative hierarchy, which allows us to capture disassortative structures only to some extent, i.e., the off-diagonal blocks can have a higher density  than the diagonal blocks. 
But the assortative hierarchy may fail to capture the community structure when the distinction between resolutions is contained in the off-diagonal, e.g., \Cref{fig:schematic_hier}B in which the diagonal blocks are homogeneous. 
A common example of networks of this type are bipartite networks in which the diagonal blocks contain no edges. 
A more general hierarchical structure may be constructed, as depicted in Figure~\ref{fig:schematic_hier}C, by combining both assortative and disassortive hierarchical features.

\subsection{Stochastic externally equitable partitions}
\begin{figure}
    \centering
    \includegraphics[width=\columnwidth]{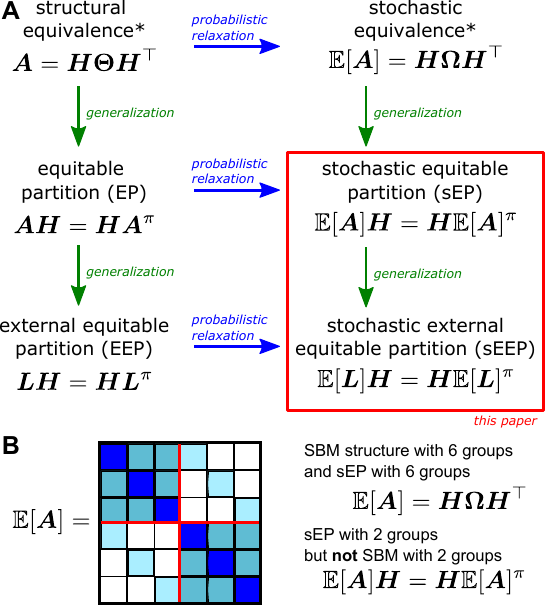}
    \caption{\textbf{Overview of network partition equivalence relationships.}
        (A) The left column describes partitions (represented by a partition indicator matrix $\bm H$) into groups of equivalent nodes in a given graph (represented by an adjacency matrix $\bm A$). 
    The right column presents the corresponding probabilistic relaxation in which the equivalence relation is considered in terms of the expected adjacency matrix $\mathbb{E}[\bm A]$ over the ensemble of networks generated by a random graph model
    ($^*$ Note that for simplicity we allow for graphs with self-loops in the algebraic expressions of structural and stochastic equivalence).
        \emph{Structural equivalence:} Nodes are equivalent if they link to the same neighbors. Here $\bm \Theta$ is a $\{0,1\}$ matrix. 
        \emph{Stochastic equivalence:} Nodes are structurally equivalent in expectation. 
        \emph{Equitable partition:} Nodes are equivalent if they have the same number of links to equivalent nodes.
        \emph{Stochastic equitable partition:} The partition is an EP \emph{in expectation}. 
        \emph{Externally equitable partition:} Nodes are equivalent if they have the same number of links to equivalent nodes, outside their own group.
        \emph{Stochastic externally equitable partition:} The partition is an EEP \emph{in expectation}.
        (B) Example of a network model with a partition into 6 and 2 groups.
        The partition into 6 groups is consistent with an SBM and an sEP.
        The partition into 2 groups is consistent with an sEP (and an sEEP) but not an SBM because the link probabilities are not uniform within the blocks.
    }%
    \label{fig:equivalences}
\end{figure}

To capture these types of generalized hierarchies, we define hierarchical communities by introducing the concept of stochastic externally equitable partitions, and describe their relationship to the stochastic block model. \Cref{fig:equivalences} provides an overview of relevant concepts and equivalence relations and how they relate to each other. 

For a given set of parameters, the SBM provides a parametric probability distribution over adjacency matrices. 
The expected adjacency matrix of this distribution can be calculated from the affinity matrix $\bm{\Omega}$ and group indicator matrix $\bm{H}$,
\begin{equation}
    \mathbb{E} [\bm A]  = \bm H \bm{\Omega} \bm H^\top \enspace .
    \label{eq:equit}
\end{equation}

The expected adjacency matrix $\mathbb{E} [\bm A]$ induces a connected weighted graph, in which all nodes in the same group are associated with exactly the same pattern of weighted edges.
In the network described by $\mathbb{E} [\bm A]$, nodes in the same group are therefore \textit{structurally equivalent}~\cite{lorrain1971structural} as they have the exact same set of neighbors and the same set of edge weights. 
In a network, represented by an adjacency matrix $\bm A$ generated from the SBM, nodes in the same group are \emph{stochastically equivalent} as they connect to the  rest of the nodes in the network according to the same set of probabilities (which are precisely $\mathbb{E} [\bm A]$). 
Put differently, groups of nodes in a network that share the exact same  set of connections are structurally equivalent. 
When groups of nodes share the exact same  set of connections \textit{in expectation} then they are stochastically equivalent.  
In this way we can consider stochastic equivalence as a probabilistic relaxation of structural equivalence (Fig.~\ref{fig:equivalences}A top row).

When we partition an adjacency matrix $\bm A$ such that every node in a group $r$ has simply the same number of links to nodes in group $s$, then we call such a partition of a graph an \textit{equitable partition}~\cite{godsil2013algebraic}.
Equitable partitions are a generalization of structural equivalence in which each node in the same group has the same sum of weights connecting it to every other group. 
(Note that here we will use the convention that the number of links, or \textit{degree}, of a node refers to the sum of edge weights when the graph is weighted.)
However, it is not necessary that nodes in the same group have exactly the same connections. 
Equitable partitions are closely related, but not identical to, graph automorphism groups~\cite{godsil2013algebraic,kudose2009equitable}, and regular equivalence~\cite{white1983graph,brandes2004structural}. 
Regular equivalence, for instance, does not require equivalent nodes to have the \emph{same number of links} to equivalent nodes, whereas equitable partitions do have this requirement.

We can extend the concept of equitable partitions to random graph models by introducing a probabilistic relaxation, which we will call a \textit{stochastic equitable partition} (Fig.~\ref{fig:equivalences}A middle row).
Partitioning the expected adjacency matrix $\mathbb{E}[\bm A]$ according to $\bm H$ creates a stochastic equitable partition such that every node in group $r$ has the same expected number of links to nodes in group $s$. 

We can define equitable partitions algebraically using an aggregated graph with adjacency matrix $\bm{A}^g \in \mathbb{R}^{k\times k}$ in which each node represents a group and the weighted links indicate the sum of link weights between groups in a graph $\bm A$:
\begin{equation}
    \bm{A}^g = \bm H^\top \bm A \bm H \enspace .
  \label{eq:agggraph}
\end{equation}
However, since groups may be of different sizes it is often more practical to use the quotient graph with weighted adjacency matrix $\bm{A}^\pi$ in which the aggregated graph $\bm A^g$ is normalized by the size of the groups:
\begin{align}
  \bm{A}^\pi 	&= \bm N^{-1} \bm H^\top \bm A \bm H 
    		= \bm H^\dagger \bm A \bm H \enspace ,
  \label{eq:quotgraph}
\end{align}
where $\bm N := \bm H^\top \bm H$ is a diagonal matrix in which $N_{rr}$ is the number of nodes in group $r$ and $\bm H^\dagger := \bm N^{-1} \bm H^\top$ is the Moore-Penrose pseudoinverse of $\bm H$. 
 Then each element of the adjacency matrix of the quotient graph $\bm A^\pi_{rs}$ tells us the mean number of edges connecting a node in group $r$ to nodes in group $s$. 
When $\bm H$ represents an equitable partition of $\bm A$ the value $\bm A^\pi_{rs}$ is the actual number of links that every node in group $r$ has with nodes of group $s$, i.e., we have the following algebraic relation:
\begin{equation}
    \bm A \bm H = \bm H \bm A^\pi \qquad \text{for all} \quad \bm H \in \mathcal{H}_{\rm EP}^{A} \enspace ,
  \label{eq:quotcond}
\end{equation}
where $\mathcal{H}_{\rm EP}^{A}$ is the set of equitable partitions of $\bm A$.

When we consider partitions that are equitable only between different groups $r\neq s$, then the partition is called an \textit{externally equitable partition} (EEP). 
We can characterize EEPs algebraically by following Eqs.~\eqref{eq:agggraph}--\eqref{eq:quotcond} and substituting the combinatorial graph Laplacian $\bm L = \bm D - \bm A$ in place of the adjacency matrix~\cite{Schaub2016}, where $\bm D =\text{diag}(\bm A \bm 1)$ is a diagonal matrix of degrees. 
This substitution gives:
\begin{equation}\label{eq:EEPalgebraic}
  \bm L \bm H = \bm H \bm L^\pi \qquad \bm H \in \mathcal{H}_{\rm EEP}^{A} \enspace,
\end{equation}
where $\mathcal{H}_{\rm EEP}^{A}$ is the set of external equitable partitions of $\bm A$, $\bm L^\pi$ is the Laplacian of the quotient graph,
\begin{align}
  \bm{L}^\pi 	&= \bm N^{-1} \bm H^\top (\bm D - \bm A) \bm H \\
    		&= \bm D^\pi - \bm A^\pi \enspace ,
  \label{eq:quotlap}
\end{align}
and $\bm D^\pi = \text{diag}(\bm A^\pi \bm 1)$ is the diagonal matrix of node degrees by group. 
Substituting the Laplacian for the adjacency matrix enables us to ignore the internal connectivity and only constrain the external connections to be equitable.
The reason that the quotient Laplacian ignores the internal connectivity is its invariance under the addition of edges in the diagonal blocks of the adjacency matrix $\bm A$, as the following proposition illustrates. 
\begin{proposition}
Let $\bm H$ be the indicator matrix of an EEP and $\bm A' = \bm A + \Delta\bm A$ be an adjacency matrix with additional within-group edges, i.e., edges that occur within the diagonal blocks.
Then $\bm{L}^\pi(\bm A') = \bm{L}^\pi (\bm A)$.
\begin{proof}
\begin{align}
    \bm{L}^\pi(\bm A') 	&= \bm N^{-1} \bm H^\top (\bm D' - \bm A') \bm H \notag\\
                     	&= \bm N^{-1} \bm H^\top (\bm D  - \bm A + \Delta\bm D- \Delta\bm A ) \bm H \notag \\
                        &= \bm D^\pi - \bm A^\pi + \bm 0 =\bm{L}^\pi (\bm A)  \enspace ,
\label{eq_qLap_ignore_diag}
\end{align}
where $\Delta \bm D$ is the diagonal matrix $\text{diag}(\Delta \bm A\bm 1)$.
The final equality in Eq.~\eqref{eq_qLap_ignore_diag} is due to the fact that $\Delta \bm A$ only contains edges in the diagonal blocks and so the diagonal matrix $\bm H^\top \Delta \bm A'\bm H = \bm H^\top\Delta \bm D'\bm H$ is equal to the group sum of degrees. 
\end{proof}

\end{proposition}

As for the EP, we propose a probabilistic relaxation for an EEP: a \emph{stochastic externally equitable partition} (sEEP) is a partition that is externally equitable \textit{in expectation} (Fig.~\ref{fig:equivalences}A bottom row).
A stochastic EEP is precisely the type of relationship we find at each level of a simple assortative hierarchy. 
For instance, the internal structure within the block diagonal of an assortative hierarchy may be further refined, but the probability of connections within the off-diagonal blocks should be uniform.
This construction can be precisely captured by an sEEP.
As a concrete example, in Figure~\ref{fig:schematic_hier}A, both the partition $\{a,b,c,d,e\}$ and the partition $\{\{a,b\},\{c,d,e\}\}$ of the expected adjacency matrix are externally equitable. 
However, stochastic EEPs also enable us to describe the more general forms of hierarchical structure shown in Figures~\ref{fig:schematic_hier}B and C.
Specifically, in an sEEP the links between nodes inside a block do not need to be uniformly distributed, but merely the expected degree with respect to every external block has to be the same.
Together with the fact that the distribution of the parameters inside the diagonal blocks in an sEEP is flexible this constitutes the main difference from the SBM (see Figure~\ref{fig:equivalences}B). 
Specifically, in the canonical SBM all elements within a block of $\mathbb{E} [\bm A]$ have equal weight, in an sEEP all rows and all columns within a block of $\mathbb{E} [\bm A]$ sum to the same value, whereas in the microcanonical SBM~\cite{peixoto2012entropy} it is the number of edges (or sum of weights) in a block of $\bm A$ that is fixed. 
This difference allows for a more flexible modeling of hierarchies than the canonical SBM, while maintaining a conceptually well defined setup. 

We therefore use the concept of a \emph{stochastic externally equitable partition} (sEEP) as the basic building block for hierarchical modular structure in networks. 
Specifically, we say the communities of a graph are hierarchically organized, if the graph's adjacency matrix can be partitioned into a sequence of nested stochastic externally equitable partitions. 
More precisely, there is sufficient evidence for a hierarchical partition if at each level of the putative hierarchy the partition is a stochastic externally equitable partition (an EEP in expectation).

If we want to recover such a hierarchy, our goal is therefore to obtain the partitions at each of the hierarchical levels, including the number of levels and number of groups at each level. 
However, before we discuss any specific method of inference, it is important to discuss some conceptual issues we face when inferring hierarchical structure from a network. 
In particular, we need not only determine when a hierarchy exists, but also how many levels are contained within the hierarchy and in which order those levels occur. 
As we will see in the next section, it is in general not possible to identify these aspects uniquely, even if we have access to the \emph{expected} adjacency matrix.

\section{Identifiability of hierarchical configurations}
\label{sec:ambiguous_hier}
Our discussion above provides us with the necessary condition for defining a set of hierarchical partitions, i.e., that they form a nested sEEP structure. However, this condition alone is insufficient to fully define a set of hierarchical communities, as we still need to resolve issues of identifiability, which we will discuss here in this section.

Identifiability is a necessary condition to guarantee that we can recover the model parameters and the hierarchy given sufficient data.  
Models of community detection (and clustering, more generally) suffer from a certain degree of non-identifiability because the community labels are permutation invariant. 
This means that there are $k!$ ways to label the same $k$ groups. 
However, this non-identifiability does not pose any problems in practice as our interpretation of each of these $k!$ solutions is identical.
When we detect hierarchical communities, we face similar issues of identifiability. 
At any given level $u$ of the hierarchy, the labels of the $k^{(u)}$ groups are permutation invariant and, as with community detection, all possible labellings of these groups represent an identical solution. 
On top of this, we can represent a hierarchy as multiple distinct dendrograms by changing how we assign branches to hierarchical levels, the order of agglomeration and/or the number of levels.

\subsection{Assigning branches to levels}
\begin{figure}
    \centering
    \includegraphics[width=\columnwidth]{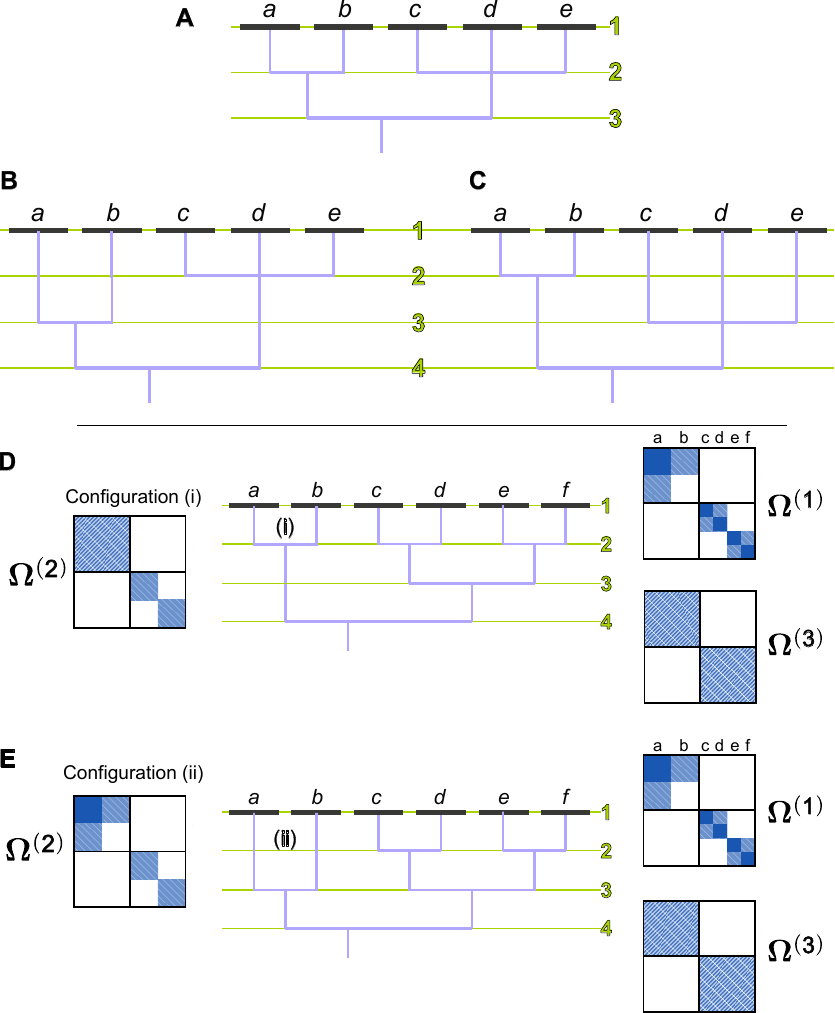}
    \caption{\textbf{Assigning dendrogram branches to levels.} 
        (A-C) Three possible assignments of hierarchical levels based on the same dendrogram. Note that (A) provides a simpler description (in terms of number of levels) and may thus be preferable.
   (D-E) Two possible hierarchical configuration for the same network. As the assignment of the dendrogram to levels has the same complexity, unless additional information is provided we cannot decide on a specific hierarchy based on the network alone. Note that the matrices $\bm{\Omega}^{(u)}$ correspond to level $u$ of the hierarchy, with groups $a$ to $f$ as indicated for $\bm{\Omega}^{(1)}$}
        \label{fig:dendro_levels}
\end{figure}

Let us assume that we already know the dendrogram structure, i.e., the rooted tree of splits of the nodes into groups, and the assignment of nodes to the groups at each branch. 
All that remains is to determine how to assign each of the branches to levels in the hierarchy. 
Figure~\ref{fig:dendro_levels} shows some examples of different ways to assign branches of a dendrogram to levels in a hierarchy.
Figure~\ref{fig:dendro_levels}A-C shows three different assignments for the same dendrogram, one assignment into three levels and two assignments into four levels. 
In each case, the main left and right branches are independent of each other and do not provide information about how we should arrange their sub-branches relative to each other. 
All three provide the same information about the hierarchical group assignment.
When confronted with equivalent solutions, a natural strategy is to take an Occam's razor approach and choose the simplest or most compact solution. 
In this case, we might therefore decide that the configuration displayed in Figure~\ref{fig:dendro_levels}A is the best choice since it has only two levels.

In other situations the ``simplest'' assignment of branches to levels may be more ambiguous. 
Figure~\ref{fig:dendro_levels}D and E show a dendrogram for which we can align the split in the left branch with either the second level (i) or the third level (ii) of the right branch. 
Both representations contain the same information about how the nodes are partitioned. However, the choice between D and E provides different aggregated affinity matrices $\bm{\Omega}^{(2)}$ (see Fig.~\ref{fig:dendro_levels}D, E) that describe how the groups of the system interact. 

It may be that in these situations a specific choice of model selection may prefer one configuration over another. 
However, as we recover the same set of groups the solutions are equivalent and therefore we should treat both solutions as the same, just as we treat partitions with different permutations of labels as being the same. 
Stated differently, the tree structure of the dendrogram remains the same, even though we interpret its branching points differently.

\subsection{Order of agglomeration}\label{ssec:agg_order}
\begin{figure}
    \centering
    \includegraphics[width=\columnwidth]{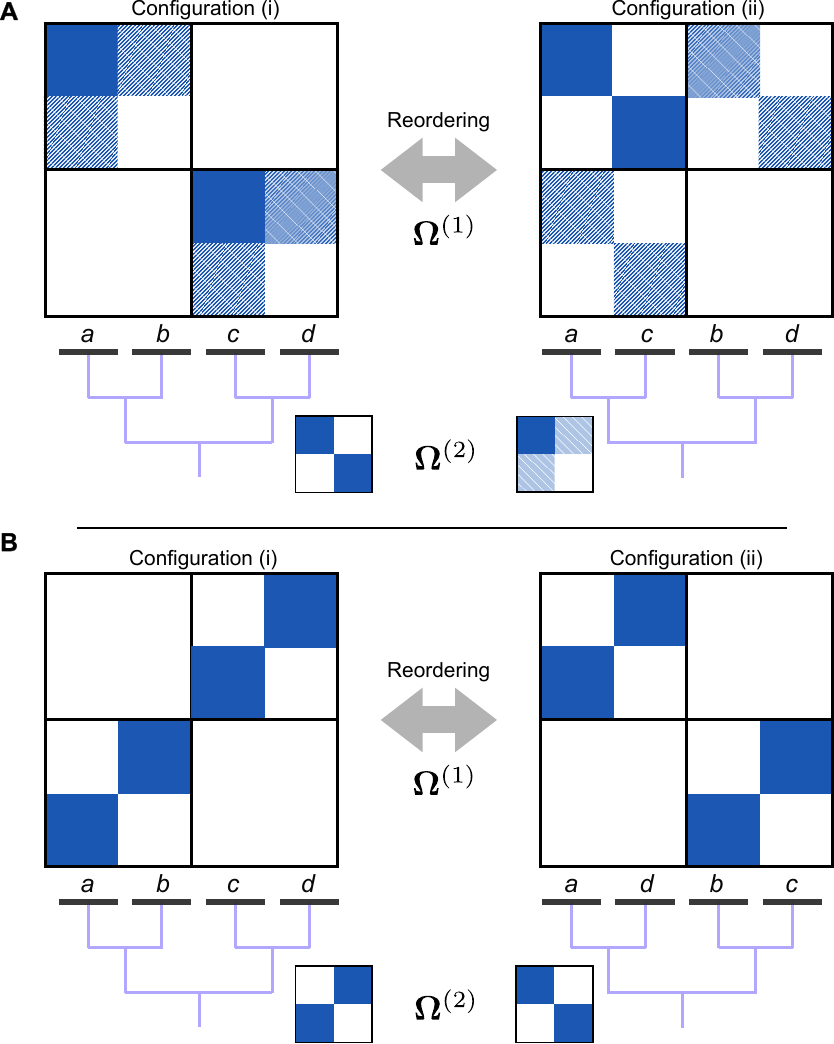}
    \caption{\textbf{Deciding the order of hierarchical agglomeration}.
        We display the pattern of two possible orderings of an affinity matrix $\bm \Omega^{(1)}$ (and its possible aggregated version $\bm \Omega^{(2)}$), indicating two possible ways of hierarchical aggregation.
        Blue represents an arbitrary link probability, white color represents zero link probability.
    (A) A~network with two possible perfect hierarchical configurations: the ordering displayed as configuration (i) may be described as two communities with inherent core-periphery structure, the ordering displayed as configuration (ii), might be thought of as a core-periphery organization with an inherent assortative modular structure.
    (B) A network with two possible hierarchical configurations, interpretable as a bipartite structure of bipartite structures; or an assortative partition of bipartite structures.}
    \label{fig:hier_config}
\end{figure}

Now consider the setting in which, instead of knowing the dendrogram, we know the desired number of layers $\ell$ in the hierarchy. 
We will also assume that we are given the partition at the finest resolution. 
All that remains is to decide which communities we should aggregate and in which order --- i.e., we want to identify the dendrogram that describes the hierarchy of communities. 
Figure~\ref{fig:hier_config} displays two example configurations for which this question is a priori ambiguous. 

Figure~\ref{fig:hier_config}A shows an example in which the parameter matrix $\bm{\Omega}^{(1)}$ of the finest level is the same for both configurations (i) and (ii), where one is just a simple permutation of the other.
However, the affinity matrices $\bm{\Omega}^{(2)}$ at the coarser level are different and so the decision of which configuration to pick depends on which version of $\bm{\Omega}^{(2)}$ we prefer. 
An appropriate form of model selection may prefer one configuration over another. 
For instance, configuration (i) has more zero blocks than configuration (ii) and so will have a higher likelihood if we assume the network was created from a nested SBM~\cite{peixoto2014hierarchical}.

The situation is different in Figure~\ref{fig:hier_config}B. Even though we have different affinity matrices $\bm{\Omega}^{(2)}$, the difference simply amounts to a different permutation of the same values and so the hierarchical configuration is non-identifiable, unless we once again include additional criteria, e.g., instead of maximising zero blocks, we might include a preference for assortative communities~\cite{ball2011efficient, gopalan2013efficient}. 
Stated differently, in both cases, had we planted one or the other hierarchy in a synthetic network, determining which hierarchy was planted would be impossible to infer from the observed data. 

\subsection{Number of levels}\label{ssec:num_levels}
Finally, let us consider the setting in which all we know is the finest partition of the network and we need to decide how to aggregate groups and how many levels there should be in the hierarchy.
Similar to the task of assigning branches to levels, it may be desirable to identify the simplest hierarchy. 
However, in this case we do not know the branches and must decide if adding levels to the hierarchy will be meaningful. 
As previously demonstrated in \cref{fig_modelvshier}, aggregating groups into any hierarchy with a particular number of levels $\ell>1$ does not imply evidence of a unique hierarchical arrangement of communities (as defined previously) in the network.
At the very least we would like to avoid including vacuous levels in the hierarchy, as in the case of \cref{fig_modelvshier}.
A clear signal of a vacuous level is a degeneracy with respect to which groups we choose to agglomerate.

As a concrete example, consider a flat partition generated from a planted partition model with $k$ groups.
In a planted partition model the affinity matrix $\bm \Omega = (a-b)\bm I + b\bm 1\bm 1^\top$ can be described with only two parameters: 
$a$, the probability that a pair of nodes in the same group will connect, and $b$, the probability that a pair of nodes in different groups will connect. 
Note that the example given in \Cref{fig_modelvshier} is a special case of the planted partition model with $k=64$, $a=1$, $b=0$. 
The partition of $\mathbb{E}[ \bm A]$ into $k$ groups will be an EEP.
If we form a new partition into $\kappa$ groups (where $\kappa < k$) by simply merging some of the $k$ groups, then the new partition will also be an EEP.
In fact \textit{any} partition formed by merging these groups will create an EEP and so every partition into $\kappa$ groups will be equivalent to each other. 
This degeneracy of partitions therefore indicates the absence of a meaningful level in the hierarchy.

\subsection{Dealing with non-identifiability and degenerate hierarchies}
As our above discussion shows, even if we had perfect knowledge about the expected adjacency matrix $\mathbb{E}[\bm A]$, uniquely identifying an underlying hierarchy is in general impossible without imposing further assumptions.
In other words, we need to impose some rules on how to break the non-identifiability issues encountered above.
In the following, we will develop a set of tools based on spectral properties associated to sEEPs, both in terms of eigenvectors as well as eigenvalues, which we will employ to detect hierarchical block structures in networks.
We emphasize that the discussion above applies generally and is not tied to any of these developments.
Specifically, our spectral approach is not the only way to resolve these issues of non-identifiability and other methods using different assumptions are conceivable as well, e.g., the already mentioned nested blockmodel by Peixoto~\cite{peixoto2014hierarchical}.

\section{Detecting hierarchies via spectral methods}\label{sec:hier_via_spectral_methods}
Thus far we have conceptualized hierarchical modular structure in terms of sequences of sEEPs based on the expected adjacency $\mathbb{E} [\bm A]$ that relates to the affinity matrix of the finest partition $\bm \Omega^{(1)}$. 
When we want to perform community detection in practice, however, we typically only have access to an observed sample adjacency matrix~$\bm A$. 
Therefore we have to either infer the precise affinity matrix $\bm \Omega^{(1)}$, which is only possible in the thermodynamic limit under certain conditions~\cite{decelle2011asymptotic}, or we have to define conditions for concluding that an sEEP is present based on the observed adjacency matrix~$\bm A$.
In combination with the identifiability issues described in the previous section the problem of detecting hierarchical communities is thus, in general, an ill-posed problem.  

In the following we will employ spectral methods to infer hierarchical community structure in a network, which correspond to a particular way of resolving the above non-identifiability issues.
Before we address these issues directly, we first outline our overall strategy to detect hierarchical community structure:
\begin{enumerate}
    \item[A.] \textit{Identify the initial finest-grained network partition.}
    	We first identify the finest level of the hierarchy (i.e., the level furthest from the root of the dendrogram) such that all nodes within a group are stochastically equivalent and ignoring the trivial partition into $n$ groups that each contain a single node. Using this initial partition we can estimate the affinity matrix of the finest partition: 
    	\begin{equation}
    	\estOmega{u} = \bm H^{\dagger(u)} \bm A \left(\bm H^{\dagger(u)} \right)^\top \enspace , 
    	\label{eq:compute_omega_agglom}
    	\end{equation} 
    	where $u=1$ (\cref{ssec:BetheHessianClustering}).   	
    \item[B.] \textit{Identify possible agglomerations and hierarchical levels.}
        Treating the estimated affinity matrix $\estOmega{u}$ as a weighted adjacency matrix, we then identify candidate partitions to form the next level in the hierarchy $\bm H^{(u+1)}$ by merging groups in the current partition such that they form an approximate sEEP (\cref{sec:sEEPgeneral}).
    \item[C.] \textit{Agglomerate and repeat.}
    	Based on the identified partitions we select the most suitable agglomeration and estimate an affinity matrix at the next level:
    	\begin{equation}
\qquad     	\estOmega{u+1} = \bm H^{\dagger(u+1)} \bm N^{(u)} \estOmega{u} \bm N^{(u)} \left(\bm H^{\dagger(u+1)} \right)^\top \enspace . \notag
		\end{equation}    
Note that $\bm H^{(u+1)}$ maps the nodes in the aggregated graph at level $u$ in the hierarchy to the nodes in level $(u+1)$ and so the dimensions of $\bm H^{(u+1)}$ will be $n^{(u)} \times n^{(u+1)}$, where the number of nodes at a given level are $n^{(u+1)} = k^{(u)}$ the number of groups at the previous level. 
We then return to the previous step and repeat until no further agglomerations are found (\cref{ssec:sEEPagglomerate})
\end{enumerate}

The key elements for addressing the non-identifiability issues are contained in steps B and C. 
First, we consider an order of agglomeration (cf. Section~\ref{ssec:agg_order}) that is induced by a singular value (or \textit{spectral}) decomposition associated with the estimated affinity matrices. 
This step may be interpreted as trying to find an agglomeration into $\kappa$ groups (where $\kappa < k^{(u)}$) that are compatible with the best rank-$\kappa$ approximation of the affinity matrix.
Second, we assess the significance of any putative agglomeration via spectral criteria to avoid inserting ``vacuous'' levels into the hierarchy (cf. Section~\ref{ssec:num_levels}). This step makes use of certain degeneracies that may exist in the spectrum, which we will discuss. 
In the next sections we explain each of the above outlined steps in detail.

\subsection{Establishing an initial partition}\label{ssec:BetheHessianClustering}
At this stage one may wonder why the identification of the initial partition is different from identifying partitions at subsequent levels in the hierarchy. 
Typically the networks we observe are sparse, meaning that the number of edges tends to scale linearly with the number of nodes $O(n)$, rather than scale according to the number of possible edges $O(n^2)$. 
In contrast, when detecting subsequent partitions we will use a (weighted) denser aggregated graph, in which nodes represent groups in the partition of the previous level. 
Different methods are better suited to sparse or dense graphs. 
In particular, sparsity is known to cause issues for detecting communities, particularly when employing spectral methods~\cite{zhang2012comparative, Krzakala2013}. 

For detecting the initial partition we will perform spectral clustering using the Bethe Hessian, which can detect communities in sparse networks right down to the theoretical limit of detectability~\cite{Saade2014}.
Furthermore, the Bethe Hessian comes equipped with a simple spectral model selection criterion that enables us to infer the number of groups~\cite{Le2019,Saade2014}.
Our experimental results confirm these theoretical studies and empirically we find that spectral clustering with the Bethe Hessian reliably identifies the finest detectable partition. 

Given the adjacency matrix $\bm A$ of a graph and the associated degree matrix $\bm D = \text{diag}(\bm A\bm 1)$, the Bethe Hessian~\cite{Saade2014} is defined as follows:
\begin{equation}\label{eq:BetheHessian}
   \bm B_\eta= (\eta^2-1)\bm I + \bm D - \eta\bm{A},
\end{equation}
where $\eta$ is a regularization parameter, which allow us to modify the spectral properties of $\bm B_\eta$ so that we can use it to detect community structure even for sparse graphs and graphs with heterogeneous degree distributions~\cite{dall2019revisiting}. 
Notice that when $\eta=1$ we recover the combinatorial graph Laplacian $\bm B_1 = \bm L$. 

Setting the regularization parameter $\eta$ to a positive value favors the discovery of assortative communities, whereas a negative value favors disassortative communities. 
As we are interested in both forms of community structure we set the regularization parameter to the positive and negative square root of the average degree $\eta = \pm\sqrt{\bm 1^\top \bm A \bm 1/n}$~\cite{Saade2014,Le2019}. 
For these settings, the number of negative eigenvalues provide a consistent estimate of the number of groups according to the SBM (see Theorem 4.3. in \cite{Le2019}).
Therefore we can use the spectral clustering with the Bethe Hessian to infer both the number of groups and the node assignments to groups at the finest hierarchical level.

We describe the exact algorithm to establish an initial partition using the Bethe Hessian in~\cref{alg:BH} in~\cref{sec:implement}.

\subsection{Identifying candidate levels in the hierarchy}
\label{sec:sEEPgeneral}
Having found an initial partition $\bm H^{(1)}$, we can estimate the affinity matrix $\estOmega{1}$ at the finest level of the hierarchy.
Treating $\estOmega{1}$ as a weighted adjacency matrix $\bm A^{(2)}$  for the second level (i.e., $\bm A^{(2)}=\estOmega{1}$), our task is now to evaluate whether or not there is sufficient evidence for a hierarchy of communities in the network.

Like other graph partitioning problems, finding all EEPs within a graph can be a computationally demanding task due to its combinatorial nature.
If we had access to the exact affinity matrix, we could adopt tools from computational group theory, which have recently shown great promise in the related problem of identifying orbit partitions within graphs~\cite{Pecora2014,Sorrentino2016,Sanchez-Garcia2018}.
However, these tools are not suitable for our task as they are only able to identify exact EEPs of the adjacency matrix, whereas we need to identify \emph{stochastic} EEPs, which are exact EEPs but only of the unobserved expected adjacency matrix $\mathbb{E}[\bm A^{(2)}] = \bm \Omega^{(1)}$.
In the best case, when a network is generated from a hierarchical model using an affinity matrix $\bm \Omega^{(1)}$ that contains a nested set of exact EEPs, our estimate $\estOmega{1} \rightarrow \bm \Omega^{(1)}$ only converges asymptotically. 
Even if we knew the true finest partition $\bm H^{(1)}$ of the generating model, statistical variation will result in minor perturbations in the estimated affinity matrix $\estOmega{1}$ relative to the true $\bm \Omega^{(1)}$.
We therefore require a new approach that enables us to define and identify stochastic EEPs within $\estOmega{1}$. 
To do so, we introduce the notion of an \textit{approximate} EEP. 
Noting that $\widehat{\bm{\Omega}} \approx \bm \Omega$, a partition that is an exact EEP of $\bm \Omega$ will be approximately an EEP of $\widehat{\bm{\Omega}}$. 
We now turn our attention to detecting approximate EEPs as a proxy for sEEPs.

\subsubsection{Finding approximate EEPs}
Central to our pursuit of identifying approximate EEPs is the fact that the presence of an (exact) EEP induces a particular structure on the eigenspaces of the Laplacian~\cite{Schaub2016,OClery2013}.
\begin{proposition}
    Let $\bm L$ be the graph Laplacian of a weighted, undirected graph with an EEP consisting of $k$ groups, described by the indicator matrix $\bm H$.
    Then, there exist $k$ eigenvectors $\bm V_k = [\bm v_{\cdot 1},\ldots,\bm v_{\cdot k}]$ and corresponding eigenvalues $[\lambda_1, \ldots, \lambda_k]$, where $\bm L\bm v_{\cdot i} = \lambda_i \bm v_{\cdot i}$, such that the values of $\bm v_{\cdot i}$ are piecewise constant for nodes within each group. 
    \label{prop_pwconst_eigenvecs}
\begin{proof}
If $\bm H$ represents an EEP and the corresponding quotient Laplacian $\bm L^\pi$ has a matrix of eigenvectors $\bm V_k^\pi = [\bm v_{\cdot 1}^\pi,\ldots,\bm v_{\cdot k}^\pi]$, then 
\begin{equation}
    \bm L \bm V_k = \bm L \bm H \bm V_k^\pi = \bm H \bm L^\pi \bm V_k^\pi =  \bm H \bm V_k^\pi \bm \Lambda^\pi = \bm V_k \bm \Lambda^\pi \enspace ,
\end{equation}
where $\bm \Lambda^\pi$ is the diagonal matrix of eigenvalues of $\bm L^\pi$.
\end{proof}
\end{proposition}

The above proposition tells us that when a network contains an EEP, then there exists a set of eigenvectors $\bm V_k$ that can be written as a linear combination of the group indicator matrix $\bm H$, i.e., there exists a matrix $\bm Q\in \mathbb{R}^{k\times k}$ such that $\bm V_k = \bm H \bm Q$.
Thus $\bm V_k = \bm H \bm V_k^\pi$ is a valid set of eigenvectors of the Laplacian $\bm L$ that are constant for nodes within the same group.
We will refer to these eigenvectors that contain this special piecewise structure as \emph{structural eigenvectors}.

For an exact EEP the variation of any structural eigenvector $\bm v$ within each group is zero. It follows then that we can characterize an approximate EEP according to the error of approximating the eigenvectors as piecewise constant.
To calculate this error, we use the matrix $\bm H^{}\bm H^\dagger$, in which $[\bm H^{}\bm H^\dagger]_{ij} = 1/n_r$ if nodes $i$ and $j$ belong to the same group $r$ and $0$ otherwise, to define a projection orthogonal to the partition $\bm H$ 
\begin{equation}\label{eq:projection_onto_H}
    \bm P_{\bm{H}} := [\bm I -\bm H^{}\bm H^\dagger] \enspace,
\end{equation}
in which $\bm H^{}\bm H^\dagger$ is used to calculate a group-wise mean such that the operator $\bm P_{\bm{H}}$  computes the matrix of residuals.
Then we can calculate the \emph{squared projection error} using the Frobenius norm $||\cdot||_\F$:
\begin{equation}\label{eq:proj_error}
    \varepsilon(\bm H, \bm V_k) :=  \|\bm{P}_{\bm{H}} \bm V_k \|_\F^2 \enspace .
\end{equation}

In~\cref{sec:sEEPconsistency} we provide evidence that minimizing this projection error is consistent with finding approximate EEPs. Consequently we can search for an approximate EEP $\widehat{\bm H} $ by minimizing the projection error:
\begin{equation}\label{eq:opt_projection_error}
    \widehat{\bm H} = \arg \min_{\bm H \in \mathcal{H}_k} \left\|\bm{P}_{\bm{H}} \bm V_k \right\|_\F^2 \enspace ,
\end{equation}
where $\mathcal{H}_k$ is the set of all partition indicators matrices with $k$ non-empty groups. 

Geometrically, the above optimization problem amounts to finding $k$ group-indicator vectors in an $n$-dimensional space, such that the $k$ vectors $\bm V_k$ will have the smallest possible variation within each group (i.e., they will be approximately constant in each group). 
Interestingly, rather than having to devise a new optimization algorithm for the above problem, we can solve the above problem using $k$-means to cluster the rows of the matrix~$\bm V_k$. We provide this proof in~\cref{sec:kmeans_proof}.

Since there exist well developed algorithms to solve the $k$-means problem this duality enables us to efficiently search for a candidate EEP when given a set of putative structural eigenvectors. 
In particular, there exist algorithms that can provide us with a provable $(1+\delta)$ approximation of the true solution of the $k$-means problem~\cite{Kumar2004}.

Connections between spectral clustering of graphs and $k$-means have previously been reported in the literature (see, e.g.,~\cite{Dhillon2004}), but only in relation to simple \textit{assortative} clusters. 
The duality we present here shows that the $k$-means procedure, when applied to the relevant eigenvectors of the Laplacian, is also related to the identification of more general EEP structures, both assortative and disassortative.

\subsubsection{Selecting relevant eigenvectors}\label{sssec:eigenvector_selection}
We have established that if a network contains an approximate EEP then we can use $k$-means with a relevant set of $k$ eigenvectors $\bm V_k$ to identify the partition. 
In principle we could search all possible combinations of $k$ eigenvectors to determine the relevant set, but this approach becomes increasingly inefficient as the network size increases.

The usual approach to selecting relevant eigenvectors for spectral clustering is to choose the eigenvectors associated with the first $k$ eigenvalues~\cite{von2007tutorial}, where ``first'' refers to either the smallest or largest values (either in terms of the real or absolute value) depending on the specific operator used. 
If we take this approach using the combinatorial Laplacian $\bm L$ then we would be constrained to identify either only assortative groups (if we use the lowest) or only disassortative groups (if we use the highest). 
In order to detect both assortative and disassortative groups at the same time, we propose to use the eigenvectors associated to eigenvalues with the $k$ largest absolute values of the \emph{uniform random walk transition matrix} $\bm W$:
\begin{equation}\label{eq:uniform_rw}
    \bm W = \bm I - \frac{1}{d_\text{max}} \bm L \enspace , 
\end{equation}
where $d_\text{max} = \max_i(D_{ii})$ is the maximal weighted degree of any node in the graph. 
Notice that $\bm W$ is simply a shifted and scaled version of the Laplacian, which has previouly been considered in the analysis
of consensus dynamics and distributed averaging~\cite{olfati2007consensus}.

The matrix $\bm W$ is a doubly stochastic matrix that describes a diffusion process on the network. Specifically, a diffusion process that has a uniform stationary distribution such that all nodes are visited with equal probability. 
Importantly, $\bm W$ has the same eigenvectors as $\bm L$ and so the aforementioned desirable spectral properties of $\bm L$ also apply to $\bm W$. 
The difference is that the set of eigenvalues $\{\lambda_i\}$ of $\bm W$ are normalized such that $\lambda_i \in [-1,1]$. 
Eigenvectors associated with positive eigenvalues correspond to assortative partitions. 
The eigenvector associated with the largest possible positive eigenvalue $\lambda_i =1$ is the vector of ones $\bm 1$ and groups all nodes into a single group (assuming the network comprises a single connected component). 
Eigenvectors associated with negative eigenvalues correspond to disassortative partitions, where an eigenvector associated with eigenvalue $\lambda_i =-1$ will describe a bipartite split in a network with a uniform degree distribution, if such a partition is possible.
Choosing the eigenvectors of $\bm W$ associated with the $k$ eigenvalues with the largest magnitude therefore allows us to detect both assortative and disassortative groups.

Note that the choosing the top $k$ eigenvectors according to absolute value may equivalently be interpreted in terms of choosing the top $k$ singular values and associated singular vectors of the matrix $\bm W$, i.e., performing the best possible rank-$k$ of $\bm W$.
Rewriting Eq.~\eqref{eq:uniform_rw} in terms of the affinity matrix $\bm \Omega$ and its degree matrix $\bm D_{\bm \Omega}=\text{diag}(\bm \Omega \bm 1)$:
\begin{equation}
    \bm W(\bm \Omega) = \bm I - \frac{1}{d_\text{max}} \bm D_{\bm \Omega} + \frac{1}{d_\text{max}}\bm \Omega \enspace , 
\end{equation}
we see that our choice of eigenvectors corresponds essentially to performing a rank-$k$ approximation of the affinity matrix, i.e., we try to find partitions into $k$ groups that best approximate the (rescaled and shifted) affinity matrix.

\subsection{Assembling the hierarchy}\label{ssec:sEEPagglomerate}
We have described an approach to detect approximate EEPs with a prescribed number of groups within an estimated affinity matrix~\estOmega{i}. 
We now describe how we can use this approach to detect and construct a hierarchy of communities from a network. 
Specifically, in the following we discuss how to determine if a partition into $k$ groups is significant enough to be included in the hierarchy and how we can identify degeneracies to avoid constructing misleading hierarchies.

\subsubsection{Assessing the significance of approximate EEPs}\label{ssec:sEEPsignificance}
Using the duality between $k$-means clustering and minimizing the projection error we can efficiently search for the partition closest to an EEP given a set of $k$ eigenvectors $\bm V_k$.
Optimizing~\cref{eq:opt_projection_error} via $k$-means will however always provide a result, even if the inferred partition $\widehat{\bm H}$ is far from being an EEP.
Therefore, it is necessary to check if the resulting partition $\widehat{\bm H}$ is significantly close to being an EEP.

We test for significance by comparing the projection error $\varepsilon(\widehat{\bm H},\bm V_k)$ against the expected projection error $\mathbb{E} \left [ \varepsilon(\widehat{\bm H},\bm U)\right ]$ under the null hypothesis that the set of eigenvectors is sampled uniformly at random from the set of all orthogonal matrices $\bm U \in \mathbb{R}^{n\times k}$, i.e.,~those matrices $\bm U$ for which $\bm U^\top \bm U = \bm I$. 

To see how we can calculate this expectation, let us start by examining a random matrix $\bm U \in \mathbb{R}^{n \times k}$ of $k$ orthonormal vectors of dimension $n$.
The squared Frobenius norm $\|\bm U \|_\F^2 = \text{trace}(\bm U^\top \bm U)$ of such a matrix will be equal to $k$.
We can compute the expectation of the square of each individual entry of $\bm U$ as:
\begin{align}
    k &=  \E\left[\left\|\bm U\right\|_\F^2\right] = \mathbb{E}\left[\sum_{j=1}^k \sum_{i=1}^n U_{ij}^2\right]
    = \sum_{j=1}^k \sum_{i=1}^n \mathbb{E}\left[ U_{ij}^2\right]\nonumber\\
      & = kn  \cdot \mathbb{E}\left[U_{ij}^2\right] \enspace,
\end{align}
where in the last step we used the fact all of the entries $U_{ij}$ are statistically equivalent.
We can conclude by symmetry that 
\begin{equation}
 \mathbb{E}[U_{ij}^2] = 1 /n \enspace ,
\end{equation} 
for all indices $i = 1,\ldots,n$ and $j = 1,\ldots,k$. 

Now, let us consider the spectral decomposition of the projection matrix $\bm P_{\bm H}$ associated with a partition into $k$ groups: 
\begin{equation}
    \bm P_{\bm H} = \bm I - \bm H\bm H^\dagger = \bm Q \bm \Lambda \bm Q^\top \enspace ,
\end{equation}
where $\bm Q\in \mathbb{R}^{n\times n}$ is an orthogonal matrix and $\bm \Lambda$ is a diagonal matrix with $\Lambda_{ii} = 1$ for $i=1,\ldots,n-k$ and $\Lambda_{ii} =0$ otherwise.
We can then write the expected projection error in terms of the spectral decomposition:
\begin{equation}
    \varepsilon_0(k) = \E \left[ \left \|\bm P_{\bm H} \bm U \right \|_\F^2\right] = \E \left[ \left \| \bm Q \bm \Lambda \bm Q^\top \bm U \right \|_\F^2\right] \enspace .
\end{equation}
We can remove the left $\bm Q$ from this equation because it is an orthogonal matrix and so does not change the norm. 
Furthermore, as $\bm Q^\top \bm U$ is simply an orthogonal transformation of unit vectors, it will have the same distribution as $\bm U$.
We can therefore simplify the above as:
\begin{align} \label{eq:vareps0}
    \varepsilon_0(k) &= \E \left[ \left \|\bm \Lambda  \bm U \right \|_\F^2\right] = \E \left[ \sum_{i=1}^{n-k}\sum_{j=1}^k U_{ij}^2\right] \notag \\ 
               &= (n-k) k\cdot \mathbb{E}[U_{ij}^2] = \frac{(n-k)k}{n} \enspace ,
\end{align}
where we have made use of the fact that $\bm \Lambda$ simply picks out the first $n-k$ rows from $\bm U$ and then used our previously established result on $\E[U_{ij}^2]$.

The above derivation assumes that $\bm U$ and $\bm H$ are statistically independent of each other.
However, in our actual calculations the eigenvectors will correspond to dominant eigenvectors of the uniform random walk matrix.
Hence, we know that $\bm 1$ is always included in $\bm V_k$ and moreover, since $\bm H \bm 1 = \bm 1$ for any partition indicator matrix $\bm H$, we know that there is always a one-dimensional subspace shared between the subspace spanned by $\bm P_H$ and $\bm V_k$. 
As we show in~\cref{app:exp_proj_error} we thus have to adjust the expected error to:
\begin{equation}
	\mathbb{E} \left [ \varepsilon(\widehat{\bm H},\bm U)\right ] =
\mathbb{E}\left [\left\|\bm P_{\widehat{\bm H}} \bm U\right \|_\F^2 \right ] =    
    \frac{(n-k) (k-1)}{n-1} \enspace . \notag
\end{equation}
The intuition here is that we have to exclude the subspace spanned by $\bm 1$ and are now looking for the projection of a \mbox{$(k-1)$-dimensional} (rather than $k$-dimensional) subspace in an $(n-1)-(k-1) = (n-k)$-dimensional space.
Moreover, both of these subspaces are restricted to be orthogonal to $\bm 1$ and thus the degree of freedom for choosing such subspaces is reduced, resulting in the change of the denominator from $n$ to $n-1$.
In other words, our calculations have to account for the fact that we know that there is a one-dimensional EEP present in any connected graph. 
Since the expected error only depends on the number of groups $k$ and not the specific partition $\widehat{\bm H}$, we will refer to the above expected error simply as $\varepsilon_0(k)$,
\begin{equation}\label{eq:exp_proj_error_unconditional}
	\varepsilon_0(k) := \frac{(n-k)(k-1)}{n-1} \enspace .
\end{equation}

The error $\varepsilon_0(k)$ above is a good null model to test the hypothesis that a single approximate EEP exists in the network because it is the expected error when there are no approximate EEPs in the network.
However we are  ultimately interested in detecting hierarchies, i.e., nested sequences of approximate EEPs, so we need to create an alternative hypothesis that accounts for the presence of other potential EEPs in the network. 
Specifically, if we would like to test the hypothesis that the eigenvectors $\bm V_k$ include a subset of $\kappa$ eigenvectors of a coarser-grained approximate EEP into $\kappa$ groups (i.e., $\kappa<k$). 
Then we calculate the expected error $\varepsilon_{0}(k|\kappa)$ conditioned on an existing EEP into $\kappa$ groups as (see~\cref{app:exp_proj_error} for details):
\begin{equation}\label{eq:exp_proj_error_conditional}
    \varepsilon_{0}(k|\kappa) =
    \begin{cases}
     \frac{(n-k)(k-\kappa)}{n-\kappa} \quad & \textrm{if} \quad \kappa \leq k \leq n\\
     \frac{(\kappa-k)(k-1)}{\kappa-1} \quad & \textrm{if}\quad 1 \leq k \leq \kappa
     \end{cases} \enspace .
\end{equation}

\begin{figure*}
    \centering
    \includegraphics[width=\textwidth]{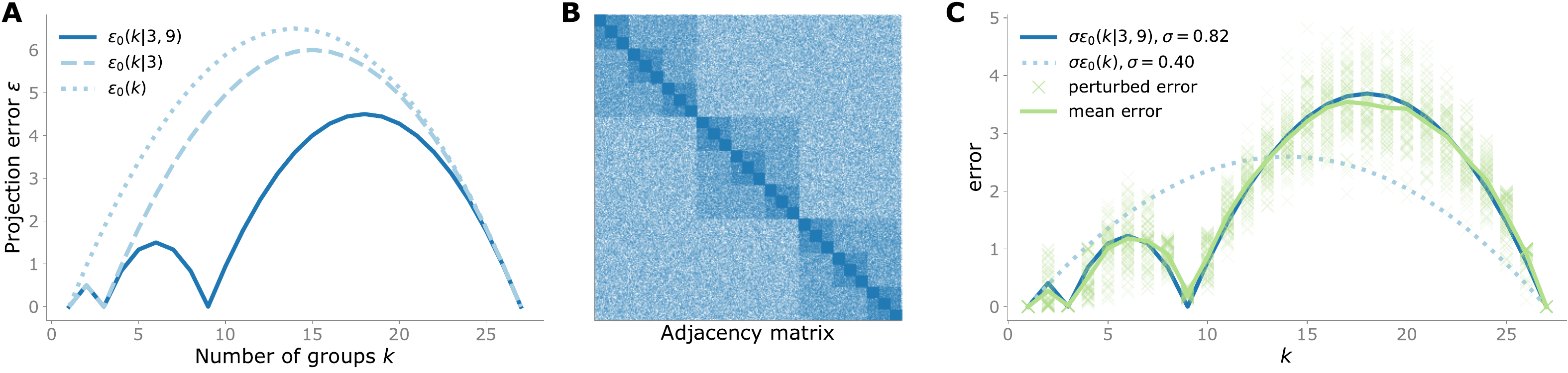}
    \caption{\textbf{Identifying levels in the hierarchy.} (A) The expected projection error $\varepsilon_0(k)$ [Eq.~\eqref{eq:exp_proj_error_unconditional}] assuming there are no hierarchical levels and the conditional expected errors $\varepsilon_0(k|\kappa)$ for $\kappa = \{3\},\{3,9\}$ [Eq.~\eqref{eq:exp_proj_error_conditional}]. (B) A spy plot of the adjacency matrix for a network with hierarchical partitions into $3, 9$ and $27$ groups. (C) Comparison of mean perturbed error against the expected error of the null hypothesis (no EEPs) $\sigma\varepsilon_0(k)$ and an alternative hypothesis $\sigma\varepsilon_0(k|3, 9)$ (EEPs into 3 groups and 9 groups). We set $\sigma$ in each case to minimize the mean squared logistic error~[Eq.\eqref{eq:MSLE}]. We clearly see the correspondence between mean error and $\varepsilon_0(k|3, 9)$. Crosses indicate the distribution of projection errors for $10^2$ random perturbations [Eq.~\eqref{eq_perturb_omega}].}%
    \label{fig:model_select}
\end{figure*}

\Cref{fig:model_select}A illustrates these expected error functions. 
The expected error $\varepsilon_0(k)$ for when there are no further approximate EEPs is shown as the dotted parabola.
However, if there is clear evidence for other levels in the hierarchy, then we need to adjust our expected error to account for these.
For example, the network represented by the spy plot in \Cref{fig:model_select}B has hierarchical partitions into $3, 9$ and $27$ groups.
To account for these possible levels we can calculate the conditional expected errors $\varepsilon_0(k|\kappa_1 =3)$ and $\varepsilon_0(k|\kappa_1=3, \kappa_2=9)$ shown in Fig.~\ref{fig:model_select}A, according to the general formula:
\begin{equation}\label{eq:exp_proj_error_cond_general}
    \varepsilon_{0}(k|\kappa_1,\ldots,\kappa_c) =
    \begin{cases}
     \frac{(n-k)(k-\kappa_c)}{n-\kappa_c} \quad & \textrm{if } \kappa_c \leq k \leq n,\\
     \qquad \vdots \\
     \frac{(\kappa_2-k)(k-\kappa_{1})}{\kappa_2-\kappa_{1}} & \textrm{if } \kappa_{1} \leq k \leq \kappa_2,\\
     \frac{(\kappa_1-k)(k-1)}{\kappa_1-1} \quad & \textrm{if } 1 \leq k \leq \kappa_1,
     \end{cases} 
\end{equation}
which can be derived analogously to~\cref{eq:exp_proj_error_conditional} for the general case.

We can decide if our candidate EEP $\widehat{\bm H}$ is significant by comparing the expected error without EEPs $\varepsilon_0(k)$ and with EEPs $\varepsilon_0(k|\kappa)$ with the observed error. 
However, before we do so, we must take precautions to prevent detecting degenerate hierarchies.

\subsubsection{Spectral signatures of degenerate EEPs and hierarchies}
By comparing the observed projection error for each putative partition using the above derived formulas, we can assess whether or not a partition is significantly close to being an sEEP.
However, as stated previously, we want to avoid constructing degenerate hierarchies, and thus we do not want to accept all possible sEEP as new hierarchical levels.

To see how this can be done, let us return to the example of a flat, non-hierarchical partition generated from a planted partition model.
After we found the split into $k$ groups, we treat the affinity matrix ${\bm \Omega = (a-b)\bm I + b\bm 1\bm 1^\top}$, as a $k \times k$ weighted adjacency matrix. 
The corresponding Laplacian is $\bm L(\bm \Omega) =  (k-1)b \bm I - b \bm 1\bm 1^\top$ and is easily identifiable as a flat partition from its spectrum:
the Laplacian $\bm L(\bm \Omega)$ has an eigenvalue $\lambda_1 = 0$, associated with the constant eigenvector $\bm 1$, and $(k-1)$ repeated eigenvalues $\lambda_r = kb$, for $2 \leq r \leq k$, associated with an invariant subspace of dimension $k-1$.
These repeated eigenvalues of $\bm L(\bm \Omega)$ clearly identify that there is no further structure in $\bm \Omega$ and there is an inherent symmetry associated to the groups.
The implication for our flat partition is that there exists a set of orthogonal matrices~$\mathcal{V}$,
\begin{equation}
    \mathcal{V} = \left\lbrace \bm V \in \mathbb{R}^{k\times (k-1)} \left| \bm V^\top \bm V= \bm I \text{ and } \bm V^\top \bm 1 = \bm 0 \right. \right\rbrace \enspace,
\end{equation}
where the columns of \emph{every} matrix in $\mathcal{V}$ form a valid set of linearly independent eigenvectors for $\bm L$. 

Consider now assessing the projection error of an EEP with indicator matrix $\bm H_\kappa$ that forms a partition on $\bm \Omega$ into $\kappa$ groups, where $1 < \kappa < k$. 
We know that there exists a matrix $\bm V_\kappa$ containing $\kappa$ dominant eigenvectors for which the projection error $\bm P_{\bm H_\kappa}\bm V_\kappa$ is exactly zero.
Based on the above observation it is easy to see that these $\kappa$ eigenvectors correspond to a particular choice of the first $\kappa$ dominant eigenvectors that are associated with one possible way to partition the network into $\kappa$ groups. 
Given that we have a flat partition, we know that \textit{any} partition into $\kappa$ groups will form an EEP and that for each partition there exists a corresponding set of $\kappa$ eigenvectors for which the projection error is zero. 
However, any given eigenvector matrix $\bm V_\kappa$ can only be piecewise constant on one of the $S(k,\kappa)$ possible EEPs into $\kappa$ groups, where the Sterling partition number $S(n, k)$ is the number of ways to partition a set of $n$ objects into $k$ non-empty subsets.
So although we can only obtain $\kappa$ independent dominant eigenvectors, there are far more possible EEPs with $\kappa$ groups, which indicates that the eigenspace is degenerate. 

The above argument can be applied analogously to situations where there are more than one level in the hierarchy and a non-identifiable set of compatible EEPs.
To capture such situations we say that an EEP into $\kappa$ groups with indicator matrix $\bm H$ is \emph{degenerate}, if some of the structural eigenvectors associated to $\bm H$ are contained within a degenerate eigenspace.
Notably, the situation here is analogous to the situation we already considered before: we are effectively picking an arbitrary subspace (corresponding to degenerate structural eigenvectors of an EEP) out of a larger degenerate eigenspace.

\subsubsection{Avoiding degenerate hierarchies}
We can use the degeneracy of EEPs to our advantage to avoid finding ``spurious'' hierarchical levels within our framework as follows. 
Recall that to find an EEP into $\kappa$ groups based on $\bm \Omega$, we consider the first dominant eigenvectors of $\bm W(\bm \Omega)$.
Now assume that the obtained EEP into $\kappa$ groups is indeed degenerate.
When we numerically compute the first dominant $\kappa$ eigenvectors, we are presented with one specific (but arbitrary) choice of eigenvectors, which depends on the specific details of the algorithm implemented. 
However, applying a small random perturbation to the affinity matrix will, with high probability, result in a different set of eigenvectors that relate to a different EEP. 
This idea also readily applies to the practical case in which we only have an estimate of the affinity matrix, $\widehat{\bm \Omega}$. The corresponding eigenspaces are only approximately degenerate since the eigenvalues will, in general, be only approximately equal.

Consider the uniform random walk matrix $\bm W$ of an estimated affinity matrix $\estOmega{u}$ and a perturbed version $\bm W_{\rm{p}} := \bm W(\estOmega{u}_{\rm{p}})$ corresponding to an affinity matrix with a slight perturbation.
We can estimate a partition $\widehat{\bm H}$ using spectral clustering on $\bm W(\estOmega{u})$.
Based on the Davis-Kahan theorem (and following an argument analogous to that in~\Cref{sec:sEEPconsistency}), we see that the difference between the eigenvectors of $\bm W$ and $\bm W_\text{p}$ will depend on how close the eigenvalues of $\bm W$ are to being degenerate.
Specifically, if the obtained eigenvectors of $\bm W_\text{p}$ and $\bm W$ are very similar, and the partition $\widehat{\bm H}$ is indeed an approximate EEP of $\bm W$, then both the projection error $\varepsilon(\widehat{\bm H},\bm V(\bm W))$ and the projection error $\varepsilon(\widehat{\bm H},\bm V(\bm W_\text{p}))$ will be small and significant (in the manner described in~\cref{ssec:sEEPsignificance}).
The robustness to small perturbations indicates that the found EEP is non-degenerate. 
However, if a small perturbation creates a $\bm W_\text{p}$ whose eigenvectors have large projection error with respect to the partition $\widehat{\bm H}$ estimated from $\bm W$, then we know that the EEP corresponds to a degenerate configuration.

In practice, once we have inferred the finest level partition into $k^{(u)}$ groups and estimated $\estOmega{u}$, for each $k_i \in \{2, ..., k^{(u)}-1\}$ we estimate a partition $\widehat{\bm H}_{k_i}$ using the $k_i$ dominant eigenvectors of $\bm W$. 
We then add a perturbation to the estimated affinity matrix such that
\begin{gather}
    \estOmega{u}_\text{p} = \estOmega{u} + \gamma \bm\Gamma \label{eq_perturb_omega} \\
    \gamma = \gamma' \frac{\|\estOmega{u}\|_2}{\|\bm\Gamma\|_2} \enspace ,
\end{gather}
where $\bm \Gamma$ is a symmetric matrix of i.i.d.~random pertubations, $\|\cdot\|$ stands for the induced $2$-norm (operator norm), and the prefactor $\gamma$ scales the perturbation of the affinity matrix to have a constant relative strength of $\gamma'$ (measured in terms of the $\ell_2$ norm).

Taking the average over perturbations gives us a mean error $\epsilon(\widehat{\bm H}_{k_i},\bm V_{k_i})$ that we can compare against the expected errors described in the previous subsection. 
We perform this comparison using the mean squared logistic error~(MSLE):
\begin{align}
	 \textrm{MSLE}(k) &= \notag \\ 
   \frac{1}{k} \sum_{k_i=1}^k & \left(
  \log \left[\epsilon(\widehat{\bm H_{k_i}},\bm V_{k_i}) + 1\right] 
  - \log \left[\sigma \varepsilon_0(k_i) + 1 \right] \right)^2 \enspace , \label{eq:MSLE}
\end{align}
where $\sigma$ is a scale parameter that we set by minimizing the MSLE. 
The MSLE is a regularized relative error that has the property of incurring a greater penalty when the expected error is small. 
This is desirable because we are more concerned with identifying the troughs, to locate approximate EEPs, than we are with matching the curvature of the peaks. 
\Cref{fig:model_select}C illustrates this comparison between the mean perturbed error and the expected error without EEPs, $\sigma\varepsilon_0(k)$ ($\sigma=0.40$), and with EEPs, $\sigma\varepsilon_0(k|3, 9)$ ($\sigma=0.82$). 
The mean error clearly is a better match with $\sigma\varepsilon_0(k|3, 9)$ indicating that there are significant EEPs into $3$ and $9$ groups. 

\subsubsection{Building a dendrogram}
Putting all of the above together, we can detect hierarchies by first identifying the finest partition and using this to estimate the affinity matrix $\estOmega{1}$, which we treat as a weighted network. 
Next we use this weighted network to identify possible partitions into $k_i \in \{2, ..., k^{(1)}-1\}$ groups and compute the corresponding projection errors (averaged over 20 perturbations) as a function of $k_i$.
We then build up a set of candidate partitions $\{\widehat{\bm H}_{\kappa_i}\}$ using a greedy heuristic. First we find the $\kappa_1$ that minimizes the MSLE between the mean perturbed projection error $\epsilon(\widehat{\bm H_{k}},\bm V_{k})$ and the expected error $\sigma\varepsilon_0(k|\kappa_1)$, i.e., 
\begin{equation}
  \arg \min_{\kappa_1} = \textrm{MSLE}(k|\kappa_1) \;, \qquad \kappa_1 \in \{2, ..., k-1\} \enspace .
\end{equation}
If $\textrm{MSLE}(k|\kappa_1) < \textrm{MSLE}(k)$ then we add $\widehat{\bm H}_{\kappa_1}$ to the set of candidate partitions. We repeat this process to add significant partitions (into $\kappa_2, \kappa_3$ etc.) until there is no further reduction in the MSLE. Note that there is no restriction on the ordering of these candidate partitions, so $\kappa_{i} > \kappa_{i+1}$ or $\kappa_{i} < \kappa_{i+1}$.
This results in a set of candidate agglomerations into $\{\kappa\}$ groups.
We pick the maximal $\max_i \{\kappa_i\}$, i.e., the finest approximate EEP to form the next level in the hierarchy and form the new affinity matrix $\estOmega{2}$.
We repeat this whole process until we no longer identify significant partitions. 

Full details of the precise algorithm are given in \Cref{sec:implement}. 
A reference python implementation of the here presented algorithms will be made available at \url{https://github.com/michaelschaub/HierarchicalCommunityDetection}.

\section{Numerical Experiments on synthetic data}\label{sec:num_experiments}

We validate the spectral algorithm introduced above for hierarchical community detection on a number of classes of synthetic networks with planted hierarchies: assortative, disassortative, symmetric and asymmetric hierarchies.  

\subsection{Experimental setup}
The synthetic network models are based on iteratively applying a planted partition model structure as follows.
We start with a planted partition model for a graph with $n$ nodes and $k=3$ groups.
We denote the probability of a link between a pair of nodes in the same group by $\alpha/n$, and denote the probability of a link between a pair of nodes in different groups by $\beta /n$. 
We set the parameters $\alpha, \beta$ by fixing an expected degree $c_1$ for each node and signal-to-noise ratio $\text{SNR}$, defined as: 
\begin{equation}\label{eq:snr_def}
    \text{SNR}(\alpha, \beta) = \frac{(\alpha-\beta)^2}{k\alpha +k (k-1) \beta} \enspace .
\end{equation}

$\text{SNR}=1$ corresponds precisely to the detectability limit of the SBM~\cite{abbe2018community,Mossel2018}. 
For each node, the expected number of connections to nodes in the same group is $a_1 = \alpha / k$, and the expected number of connections to nodes in different groups is $b_1 = \beta(k-1)/k$, such that the total expected degree for each node is $c_1 = a_1 + b_1$. 

\begin{figure}[tb!]
    \centering
    \includegraphics[]{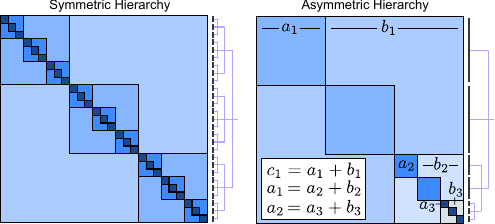}
    \caption{\textbf{Schematic: expected adjacency matrices of synthetic hierarchical test networks.}
    We consider a symmetric and an asymmetric hierarchical network construction. Both start from a planted partition model with a specified signal to noise ratio.
    We then iteratively refine the hierarchy by treating the network induced by each subnetwork as another instance of a planted partition model with the same signal to noise ratio.
    Here we impose the additional contraint that the expected degree (the average connection probability) of the nodes in this subnetwork is such that it matches with the specification of the layer above (see text for details). 
    In the symmetric variant of our model, each group is recursively sub-divided such that we obtain a hierarchy of $3\times 3\times 3 = 27$ groups.
    In the asymmetric varianot of the model, only one of the groups is subdivided further, leading highly skewed group sizes at the lowest level of the hierarchy (see the indicated dendrogram).}
    \label{fig:syn_test_networks}
\end{figure}

Next we recursively plant finer partitions, while maintaining the average node degree.
We divide each of the $k$ groups again into $k$ subgroups, such that the expected degree of the nodes in this subnetwork is $c_2 = a_1 = a_2 + b_2$, consistent with the coarser, initial planted partition.
Figure~\ref{fig:syn_test_networks} illustrates a schematic of these parameters for the symmetric and asymmetric hierarchies. 
The parameters $a_2$ and $b_2$ (respectively their connection probabilities) within each subnetwork are chosen such that the specified $\text{SNR}$ is maintained.

\begin{figure}
  \centering
  \includegraphics[width=\linewidth]{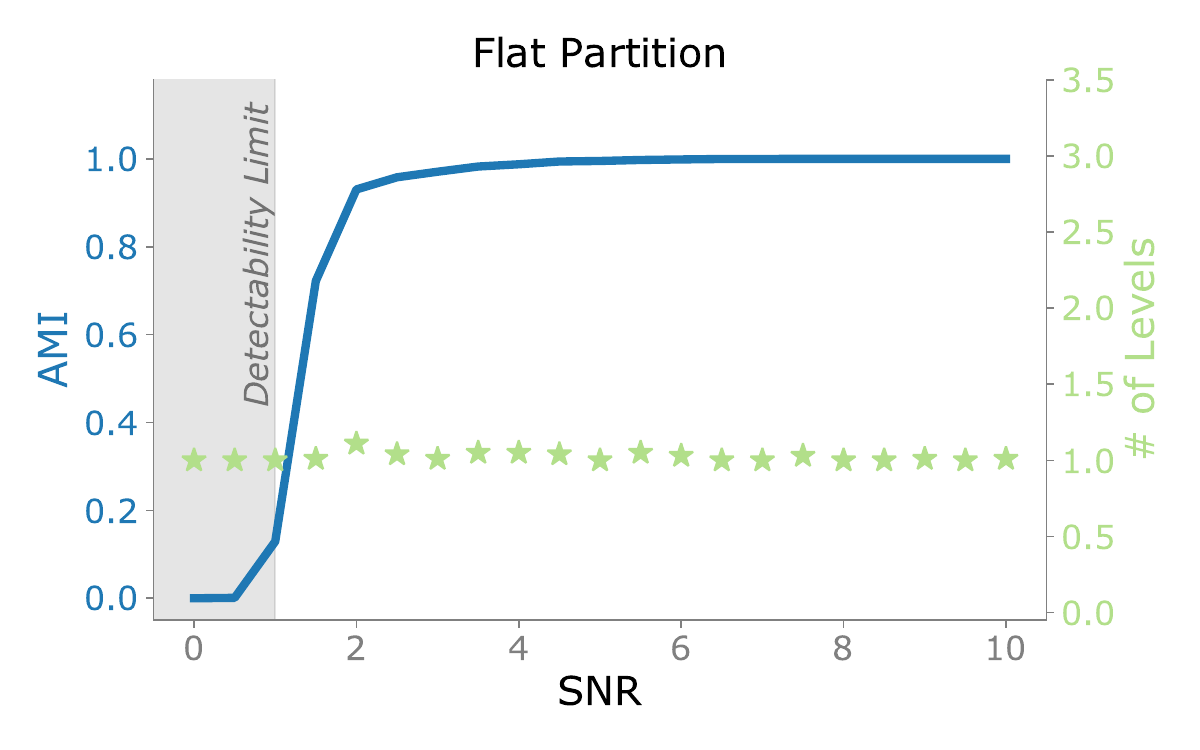}
    \caption{\textbf{Detecting the absence of hierarchy (flat partitions)}. We plant a flat partition into 64 groups, similar to the example in \cref{fig_modelvshier}. Overall our method is consistent in identifying a flat partition across the full range of SNR values.}
    \label{fig:flat_experiment}
\end{figure}

\begin{figure*}
	\centering
    \includegraphics[width=\linewidth]{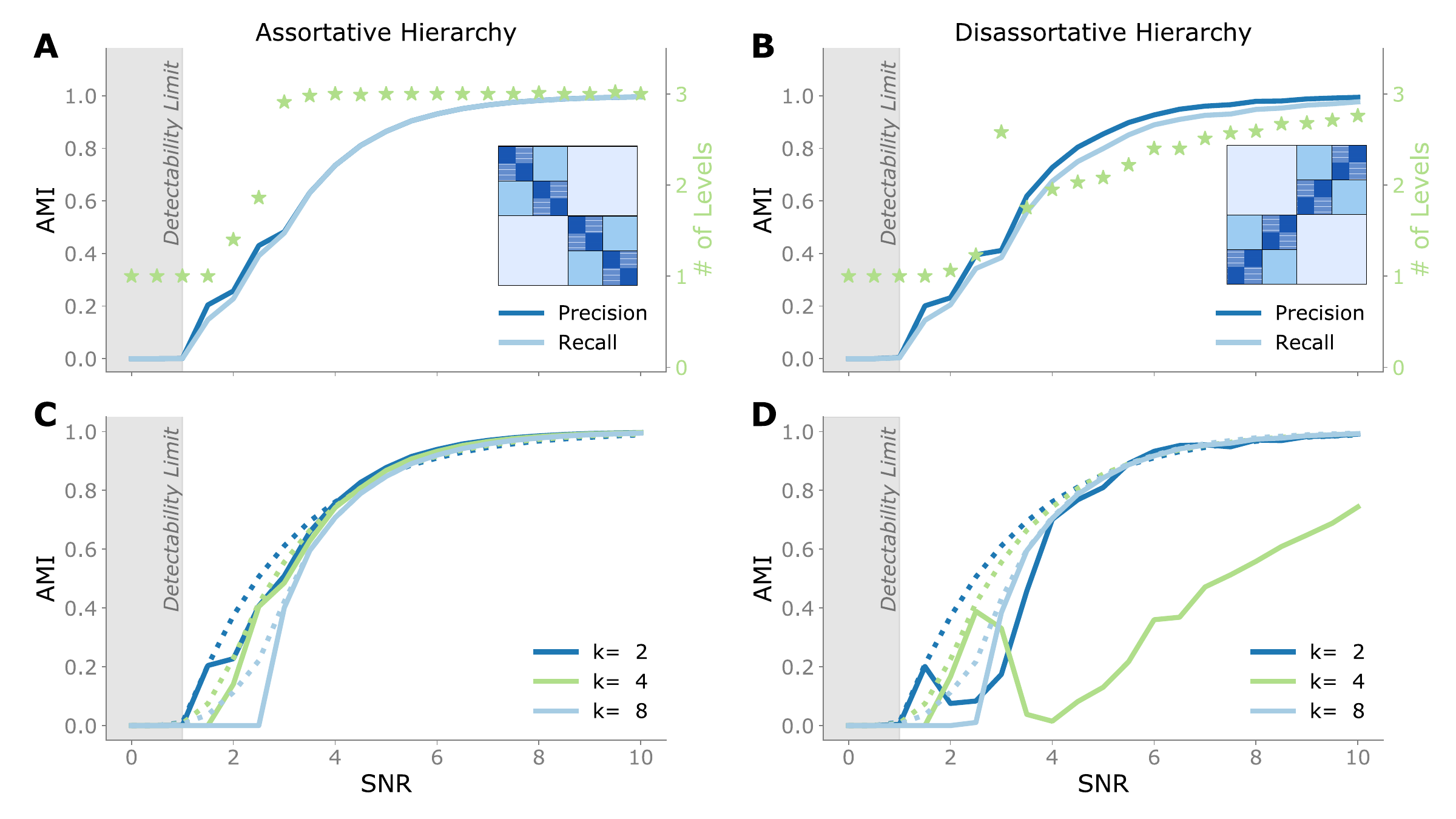} 
    \caption{\textbf{Detecting assortative and disassortative hierarchical communities planted in synthetic test networks.} 
        Synthetic networks are drawn from the assortative and disassortative hierarchical random graph model with $n=2^{14}\approx 16,400$ nodes with average degree $50$.
    Levels in the hierarchy are partitioned into $2, 4$ and $8$ groups. 
    (A) Results for the assortative hierarchical network model (see inset for schematic). 
    We show the precision and recall statistic of the overall hierarchy, as defined in the text, as a function of the signal-to-noise ratio (SNR). 
    Stars denote the number of hierarchical levels identified by our algorithm (right y axis).
    (B) The corresponding results for the disassortative hierarchical network model (see inset for schematic).
    (C) The mean Adjusted Mutual Information (AMI) of the best matching inferred partition with each level in the planted assortative hierarchy. The dotted lines indicate the performance of spectral clustering with the Bethe Hessian with known number of groups.
    (D) The mean AMI of the best matching inferred partition with each level in the planted disassortative hierarchy. 
    We observe poorer performance in recovering the disassortative hierarchy, in particular the level into $k=4$ groups because of the degeneracy of disassortative groups (see text for details). }
    \label{fig:ass_dis_exps}
\end{figure*}

\begin{figure*}
    \centering
    \includegraphics[width=\linewidth]{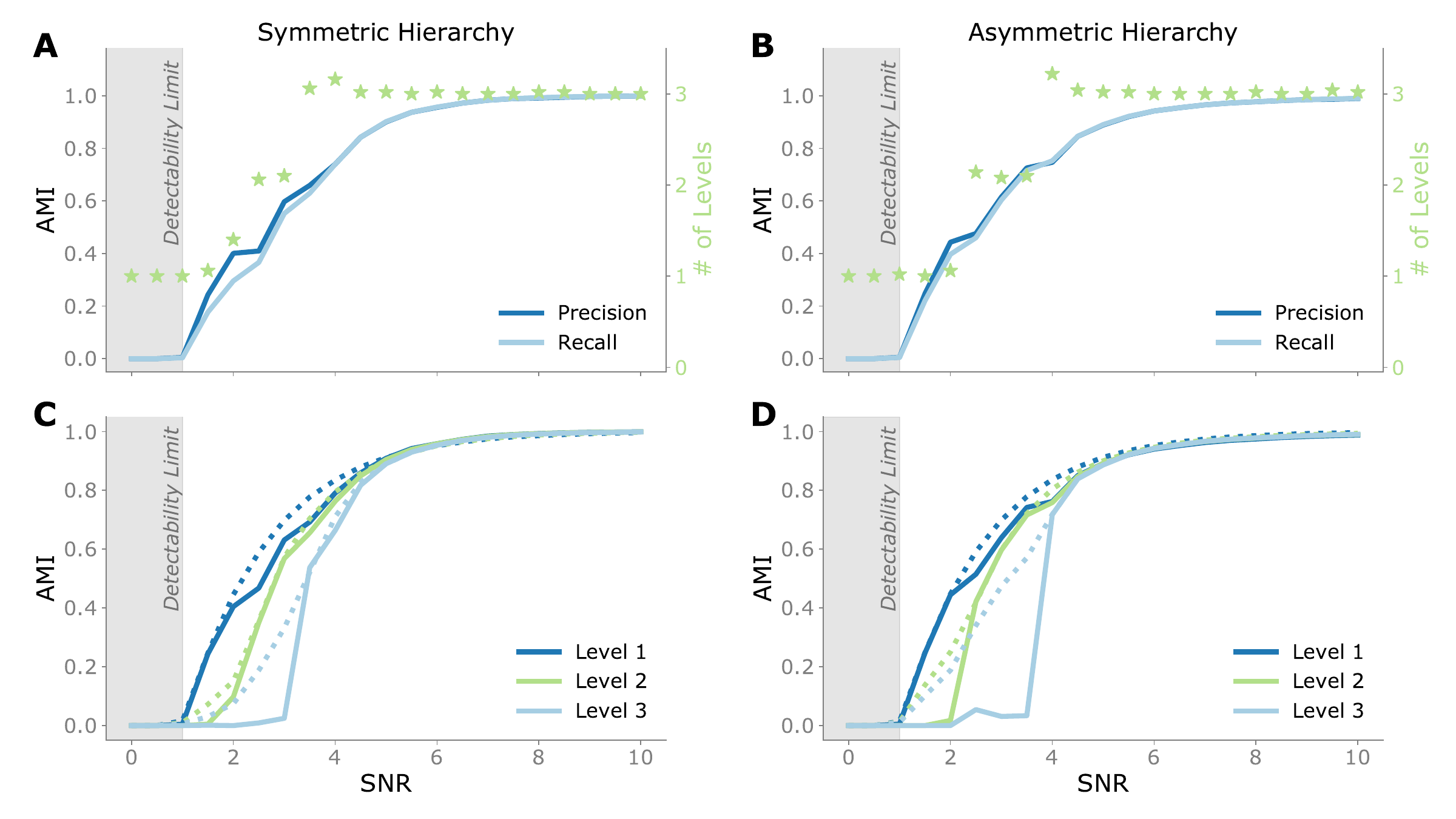} 
    \caption{\textbf{Detecting symmetric and asymmetric hierarchical communities planted in synthetic test networks.} 
    Synthetic networks drawn are from the symmetric and asymmetric hierarchical random graph model with $n=3^9\approx 19,700$ nodes with average degree $50$.
    Levels in the symmetric hierarchy are partitioned into $3, 9$ and $27$ groups, whereas the levels in the asymmetric hierarchy are partitioned into $3, 5$ and $7$ groups. 
    (A) Precision and recall for the symmetric hierarchical networks as a function of signal-to-noise ratio (SNR).  
    Stars denote the number of hierarchical levels detected by our algorithm (right y axis).\@
    (B) Precision and recall for the asymmetric hierarchical networks as a function of SNR.\@
    (C) The mean Adjusted Mutual Information (AMI) of the best matching inferred partition with each level in the planted symmetric hierarchy. The dotted lines indicate the performance of spectral clustering with the Bethe Hessian with known number of groups.
    (D) The mean AMI of the best matching inferred partition with each level in the planted asymmetric hierarchy. }%
    \label{fig:syn_experiments}
\end{figure*}

We validate our method for each class of network models for varying levels of SNR. 
To evaluate the similarity of two partitions we use the adjusted mutual information score~\cite{Vinh2010} defined as:
\begin{equation*}\small
    \text{AMI}(\bm H_1,\!\bm H_2)\!=\! \frac{I(\bm H_1,\bm H_2) - \mathbb{E}[I(\bm H_1,\bm H_2)]}{(\text{Ent}(\bm H_1) + \text{Ent}(\bm H_2))/2- \mathbb{E}[I(\bm H_1,\bm H_2)] } \enspace ,
\end{equation*}
where $I(\bm H_1,\bm H_2)$ and $\mathbb{E}[I(\bm H_1,\bm H_2)]$ are the mutual information and its expected value respectively, and $\text{Ent}(\cdot)$ is the Shannon entropy of the partition assignment.
Here the expectation is taken over the so-called permutation null model~\cite{Vinh2010}, in which partitions are generated uniformly at random subject to the constraint that the number of clusters and points in each clusters are commensurate with the inputs~\cite{Gates2017,Vinh2010}.
Note that the AMI score typically lies in the range $[0,1]$\footnote{it is possible to have slightly negative AMI values due to the adjustment for chance.} with $0$ denoting a result as expected by chance and $1$ perfect recovery.

We denote the $\ell$ planted partitions within our model networks as $\bm H_1, \ldots , \bm{H}_\ell$ and denote $\hat{\ell}$ hierarchical partitions detected by our algorithm as $\widehat{\bm H}_1,\ldots, \widehat{\bm H}_{\hat{\ell}}$. 
Using the AMI score we define the score matrix $\bm \Xi$ with entries
\begin{equation}
    \Xi_{i,j} = \text{AMI}(\bm H_i,\widehat{\bm H}_j) \text{ for } i=1,\ldots,\ell,\; j=1,\ldots,\hat{\ell} \enspace ,
\end{equation}
that measures the pairwise matching between any of the planted and recovered partitions. 
We summarize the detection performance in the score matrix using precision and recall, defined as:
\begin{align}
    \text{Precision} & = \frac{1}{\hat{\ell}} \sum_j \max_i \Xi_{i,j} \\
    \text{Recall} & = \frac{1}{\ell} \sum_i \max_j \Xi_{i,j} \enspace .
\end{align}

The precision is large if, for every estimated partition, there is a planted partition that provides a good match.
The recall is large if for every planted partition, there is an estimated partition that matches closely.

\subsection{Results}\label{ssec:syn_results}
In our first experiment we confirm that our approach does not identify degenerate hierarchies. 
We plant a flat partition into $64$ groups using a planted partition model, akin to the example in \Cref{fig_modelvshier}, and vary the SNR. 
\Cref{fig:flat_experiment} shows that our approach is broadly consistent at identifying a single partition in the absence of a hierarchy. 

Next we consider assortative and disassortative hierarchies. In both cases we generate symmetric hierarchical partitions into 2, 4 and 8 groups. We generate the disassortative hierarchies in the same way as the assortative hierarchies, except that we reverse the columns of the affinity matrix $\bm \Omega^{(1)}$ before generating the network (see insets \cref{fig:ass_dis_exps}A-B).

\Cref{fig:ass_dis_exps} shows the performance in recovering the assortative (A and C) and disassortative (B and D) hierarchies. 
In the case of the assortative hierarchy we see that the performance increases monotonically with the SNR, both overall (\cref{fig:ass_dis_exps}A) and at each level (\cref{fig:ass_dis_exps}C). 
We observe poorer overall performance in recovering the disassortative hierarchies and require a much higher SNR to consistently identify three levels in the hierarchy (\cref{fig:ass_dis_exps}B). 
Closer inspection of the performance at individual levels (\cref{fig:ass_dis_exps}D) shows that we can recover the finest partition into 8 groups using the Bethe Hessian with comparable performance as the assortative case. 
We can also detect the coarsest partition into 2 groups relatively well, particularly at $\text{SNR}>4$.
However the middle level is harder to detect. 
The reason for the poorer performance is due to a degeneracy that occurs for disassortative partitions meaning that we have multiple distinct ways to form an EEP into 4 groups~\cite{peel2020}. 
This degeneracy creates an identifiability issue, similar to the one described in \cref{fig:hier_config} (see \cref{app:identifiability} for a visual description), and means that our algorithm often fails to detect a level in the hierarchy that partitions the network into 4 groups.
Identifiability issues notwithstanding, these results indicate that our approach is still effective at recovering disassortative hierarchies.

Finally, we examine the performance of recovering symmetric versus asymmetric hierarchies. 
\Cref{fig:syn_experiments} displays the results for a symmetric hierarchy with three partitions into 3, 9 and 27 groups (\cref{fig:syn_experiments}A and C) alongside results for an asymmetric hierarchy partitioned into 3, 5 and 7 groups. 
Our algorithm shows overall good performance: not only do we recover the correct partition at the finest level, we can also detect right until the detectability limit.
The fact that the precision and recall measures are well aligned indicates that our algorithm successfully rejects spurious hierarchical levels, as can also be seen from the number of hierarchical levels found (indicated by orange asterisks in~\Cref{fig:syn_experiments}).
We detect additional levels only in a limited number of cases where the SNR increases sufficiently such that the intermediate levels become well defined.

\section{Detecting hierarchical structures in real-world data}
\label{sec:real_experiments}

To validate our method on real-world networks, we consider a face-to-face contact network and a word-association network, described in the subsequent sections.
The standard SBM has a well-known weakness for modelling real-world networks because, for network generated by the SBM, the degrees of nodes within a group are Poisson distributed~\cite{Karrer2011}. Real-world networks tend to have a more heterogeneous degree distribution, which has motivated various forms of degree correction~\cite{dasgupta2004spectral, Karrer2011}.  However, the Bethe Hessian is more robust to degree heterogeneity, but we further improve this by adjusting the regularization parameter according to Ref.~\cite{dall2019revisiting} (see Algorithm~\ref{alg:DC_BH} in Appendix~\ref{sec:implement} for details). 
Because our approach is agglomerative, where subsequent steps of the algorithm simply merge groups from the previous level, it is only necessary to account for degree heterogeneity in the initial detection of communities.

\subsection{High-School Network}
We first consider a social contact network within a high-school~\cite{mastrandrea2015contact} to identify the presence of possible hierarchical structure.
The network consists of $n=327$ nodes and $m=5818$ edges, denoting face-to-face contacts between students wearing RFID tags. 
The students are divided into nine classes according to their subject specialization: math \& physics (MP, 3 classes), biology (BIO, 3 classes), physics and chemistry (PC, 2 classes), and engineering (PSI, 1 class).

\begin{figure}
    \centering
    \includegraphics[width=\linewidth]{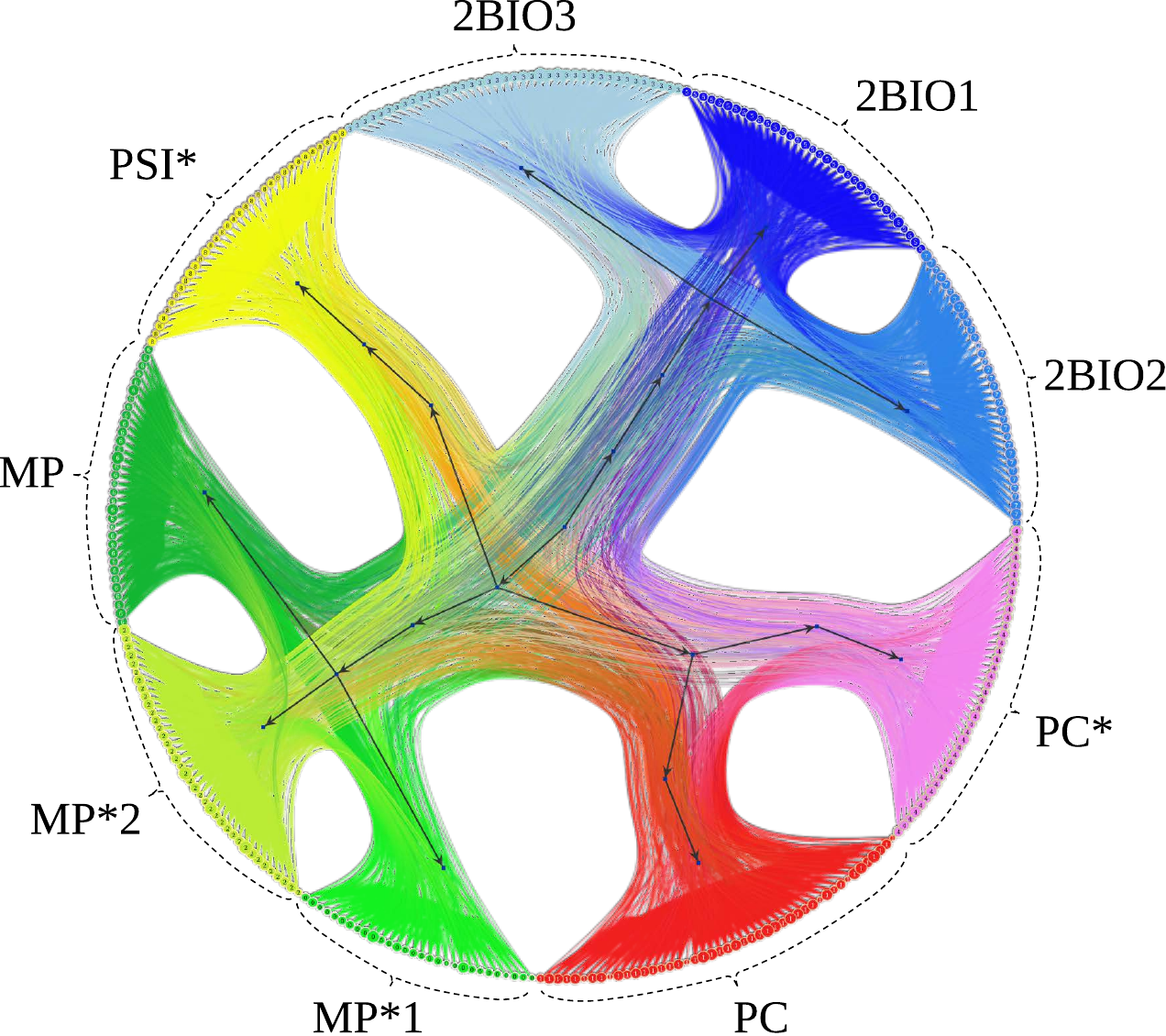} 
    \caption{\textbf{Hierarchical community structure in a high-school social contact network.}
        Using spectral clustering, we find 9 communities corresponding to the classes in the high-school network (as indicated by colors above). There are 3 classes specialized on math \& physics (MP), 3 biology focussed classes (BIO), 2 physics and chemistry focussed classes (PC), and 1 class specialized for engineering (PSI).
        As depicted above we find a hierarchical structure commensurate with these specializations, using our spectral method.
    }%
    \label{fig:highschool_hier}
\end{figure}

\Cref{fig:highschool_hier} shows the hierarchy that we identify using our spectral algorithm. 
We see that the hierarchical organization in the social contact structure of the network matches the class structure of the school.
Specifically, individual classes are  identified as individual communities at the finest level, which in turn merge with classes with the same specialization. 
Finally, the coarsest partition splits the students into two groups: those that specialize in biology and those whose specializations involve physics.

\subsection{Word associations network}
We constructed a network of English word associations using data from the The Small World of Words project~\cite{de2019small}, a scientific project to map word meaning in various languages. 
The dataset was created based on a word association task, in which participants are asked to give three associated responses to a given cue word. 
The dataset includes over 3 million responses obtained from over 90,000 participants, for more than 12,000 cues. 
We created a network of stemmed words as nodes and cue--response pairs as edges, and applied our algorithm to identify the hierarchical structure of communities. 

\begin{figure*}
    \centering
    \includegraphics[width=\linewidth]{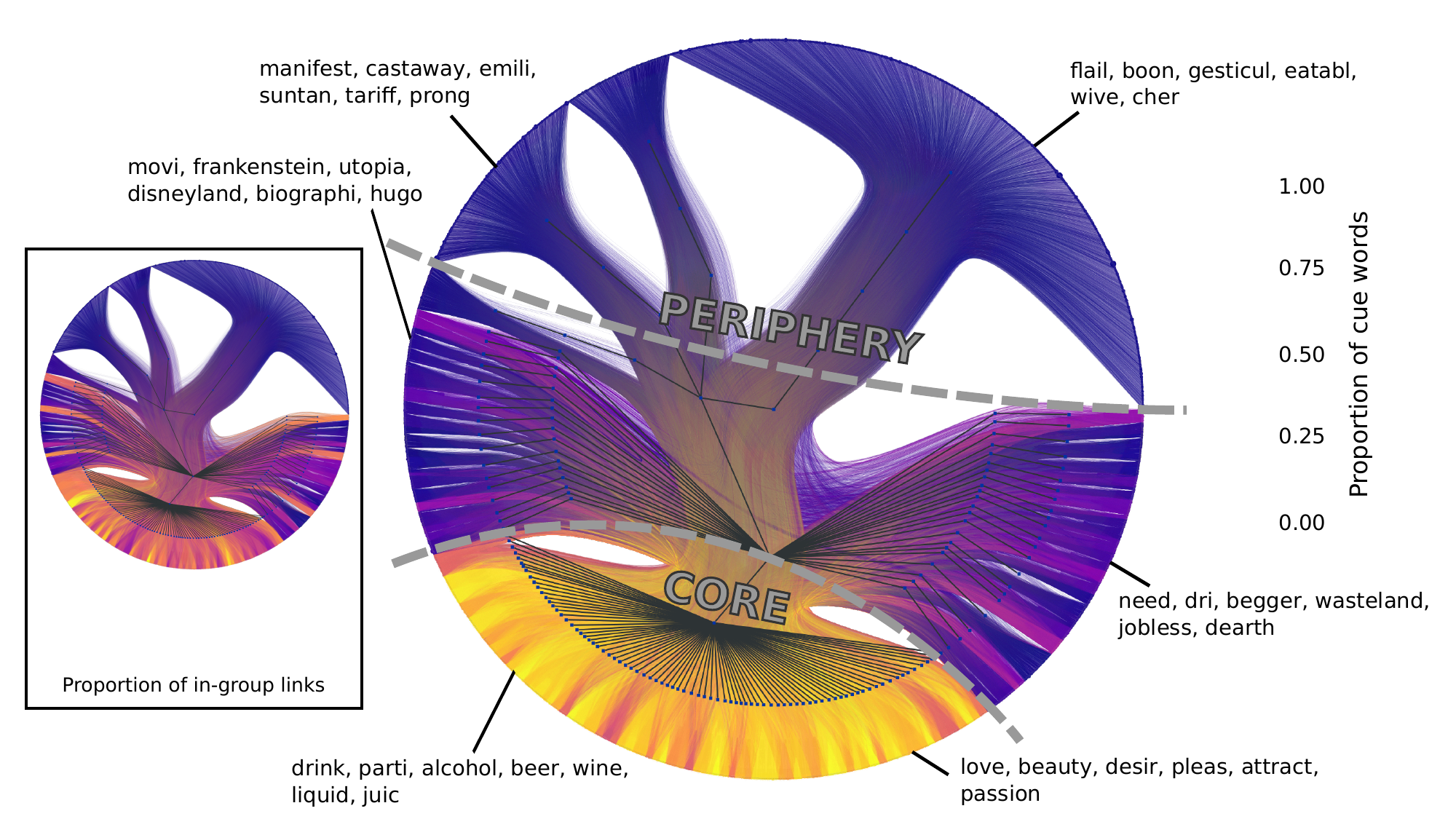} 
    \caption{\textbf{Hierarchical structure in a word association network.}
    Using our spectral method developed here, we find a hierarchical core-periphery structure in this network, as depicted above. Each node in the network corresponds to a word (either a cue or a response), and two words are connected if they form a cue-response pair. Color depicts the amount of cue-words within a group. 
    The core is formed of densely knit set of words corresponding mostly to cue words, whereas the more peripheral communities are formed mostly from response words (non-cue words).
    }%
    \label{fig:SWOW_hier}
\end{figure*}

Figure~\ref{fig:SWOW_hier} shows the  dendrogram representing the detected hierarchical structure. 
Here we see at the coarsest level  a partition into three groups that forms a core-periphery type of structure.
The nodes in the dense core have a higher proportion of in-group links and are more likely to represent a cue word.
The finer partitions of the core represent groups of words that are clearly associated, whereas the periphery contains groups of words that are less clearly associated due to the disassortative nature of the communities (i.e., lower proportion of in-group links).

\section{Conclusion}\label{sec:conclusion}
We have presented a thorough investigation on hierarchical community structure in networks.
By introducing the concept of a stochastic externally equitable partition, we have provided a formal definition of hierarchical community structure that consists of a series of nested, non-degenerate stochastic externally equitable partitions. 
Stochastic externally equitable partitions provide a natural generalization of several concepts of node equivalence. 
In particular, it has a close relationship to the stochastic equivalence relation that underlies the stochastic block model.
In light of our new definition of hierarchical community structure, we have discussed several identifiability issues that apply in general to the detection of hierarchical community structure. 
Specifically, we have identified a number of scenarios for which multiple good solutions exist.
In these cases, the choice of which hierarchy is detected will be based on the specific bias of the detection method employed. 
We have also discussed how na\"ive use of hierarchical models, such as~\cite{peixoto2014hierarchical}, may identify spurious hierarchies, in much the same way that community detection algorithms might identify spurious communities in an Erd\H{o}s-R\'enyi network. 
In addition, we have identified characteristic spectral properties of hierarchical stochastic EEPs and developed a simple, efficient algorithm for hierarchical community detection that exploits these properties.

Our work opens a number of avenues for future research. 
On a theoretical level, our work lays the foundations for more detailed analysis of the asymptotic limits of detectability of community structure, particularly for networks that contain communities at multiple resolutions, as is the case for hierarchical communities~\cite{peel2020}. 
Our experimental results further emphasize the issues of identifiability, in particular for disassortative hierarchies. 
We see that disassortative hierarchies are more likely to have degenerate solutions that make it harder to detect levels in the hierarchy and/or identify the specific planted partition over an equivalently good alternative solution. 
These observations warrant further investigation into the degeneracy of disassortative partitions, something that has been largely overlooked so far, possibly due to the bias in the literature towards assortative community structure. 
One potential solution to deal with the identifiability issues might be to incorporate a notion of equivalent hierarchies into the scoring functions we use to evaluate performance. 
We already employ a similar approach in community detection to deal with the fact that communities are invariant to their specific label assignment. 
However, this is not a consideration we have encountered so far in the body of work concerned with evaluating (hierarchical) community detection performance~\cite{gates2019element, lancichinetti2009detecting, Perotti2015}. 
From an algorithmic perspective, we have focused on an agglomerative procedure that relies on accurately detecting the finest level in the hierarchy.
Any errors in recovering the finest partition will be propagated to subsequent levels.
However, it may be that a divisive algorithm could perform better in some settings, particularly if the coarser partitions contain a stronger community structure that is easier to detect.
Investigating the relative benefits and weaknesses of agglomerative versus divisive algorithms may thus be a fruitful avenue for future research. 

\section*{Acknowledgements}
The authors would like to thank Jean-Charles Delvenne, Karel Devriendt, Mauro Faccin, Renaud Lambiotte, Tiago Peixoto, Karl Rohe, and Michael Scholkemper for helpful conversations.
MTS was partially supported by the European Union’s Horizon 2020 research and innovation programme under the Marie Sklodowska-Curie grant agreement No 702410, and by the Ministry of Culture and Science (MKW) of the German State of North Rhine-Westphalia (``NRW Rückkehrprogramm'').

\bibliography{wns-bib}
\appendix

\newpage
\section{Consistency of the projection error criterion}\label{sec:sEEPconsistency}
Central to our approach is the identification of stochastic EEPs, which satisfy:
\begin{equation}
  \bm L(\bm \Omega) \bm H = \bm H \bm L^\pi \qquad \bm H \in \mathcal{H}_{\rm EEP}^{\Omega} \enspace,
\end{equation}
where $\mathcal{H}_{\rm EEP}^{\Omega}$ is the set of external equitable partitions of $\bm \Omega$, $\bm L^\pi$ is the Laplacian of the quotient graph,
\begin{equation}
  \bm{L}^\pi = \bm N^{-1} \bm H^\top (\bm D - \bm \Omega) \bm H \enspace .
\end{equation}

Since we do not observe the affinity matrix $\bm \Omega$, we  estimate the corresponding affinity matrix \estOmegaNoSup  as:
\begin{equation}
    \estOmegaNoSup = \bm N^{-1}\bm H^\top \bm A \bm H \bm N^{-1} \in [0,1]^{k \times k} \enspace ,
\end{equation}
and assume that if $\bm H$ is an EEP of $\bm \Omega$ then $\bm H$ will be \textit{approximately} an EEP of \estOmegaNoSup, i.e., 
\begin{equation}
  \bm L(\bm \estOmegaNoSup) \bm H \approx \bm H \bm L^\pi \qquad \bm H \in \mathcal{H}_{\rm EEP}^{\Omega} \enspace .
\end{equation}
Our question now is whether or not this is a reasonable assumption and if we can use the projection error as a measure of how well a partition of \estOmegaNoSup approximates an EEP of $\bm \Omega$.
We make this argument below by demonstrating that (as $n \rightarrow \infty$) stochastic fluctuations in the realization of the adjacency matrix $\bm A \in \{0,1\}^{n \times n}$ cannot lead to a quotient graph whose Laplacian eigenvectors have a large projection error. 

Since each entry of \estOmegaNoSup correspond to a sum over independent Bernoulli random variables, by the central limit theorem each entry $\widehat{\Omega}_{ij}$ will, for large $n$, be well approximated by a Gaussian random variable $\mathcal N(\mu_{ij},\sigma_{ij})$, if the number of groups stays bounded, i.e., each group becomes sufficiently large.
The empirical mean is an unbiased estimator, so the mean of each of these Gaussians will be given by the corresponding entry of true affinity matrix $\mu_{ij} =  \Omega_{ij}$. 
Similarly, the variance of each entry will be $\sigma_{ij}^2 = \Omega_{ij}(1-\Omega_{ij}) / n_i n_j$, where $n_i$, $n_j$ are the number of nodes in group $i$ and $j$, respectively (where $\sum_j n_j = n$).
It follows that the spectral properties of \estOmegaNoSup will closely approximate the true affinity matrix $\bm{\Omega}$.
More precisely, it can be shown that the spectral norm $\|\widehat{\bm{\Omega}} - \bm{\Omega}\|_2$ will be small with high probability (see, e.g.,~\cite{Bandeira2016}).

Now, let $\bm V_\kappa$ be a matrix containing $\kappa$ structural eigenvectors (out of $k$ total eigenvectors) of $\bm L(\bm \Omega)$, where $\bm V_\kappa$ is associated with $\kappa$ consecutive eigenvalues $\lambda_i,\ldots, \lambda_j$.
Here, we have ordered the eigenvalues in ascending order such that $\lambda_1 \le \lambda_2\le \ldots\le \lambda_k$.
The Davis-Kahan Theorem~\cite{davis1970rotation} implies that the corresponding eigenvectors $\widehat{\bm V}_\kappa$ of $\bm L(\estOmegaNoSup)$ are indeed close to eigenvectors of the true quotient Laplacian $\bm L(\bm{\Omega})$ (c.f. Eq.~3 of~\cite{Yu2015}):
\begin{equation}\label{eq:DavisKahan}
    \|\widehat{\bm V}_\kappa\bm O - \bm V_\kappa \|_\F \le \frac{\sqrt{8k}\|\bm L(\estOmegaNoSup) - \bm L(\bm{\Omega})\|_2}{\Delta \lambda} \enspace ,
\end{equation}
where $\bm O$ is a unitary matrix, $\|\cdot\|_2$ is the operator norm and 
\begin{equation}
    \Delta\lambda = \min(\lambda_{i} - \lambda_{i-1}, \lambda_{j+1} - \lambda_{j})  \enspace ,
\end{equation}
is the eigenvalue gap associated with the true quotient Laplacian.
To make the above formula valid for any set of consecutive eigenvalues we define $\lambda_0 = -\infty$ and $\lambda_{k+1} = +\infty$.

As our estimated partition was such that $\bm L(\estOmegaNoSup) \approx \bm L(\bm \Omega)$, it follows that if the eigenvalue decomposition is unique and the eigenvalue gap is thus nonzero, the projection error [\cref{eq:proj_error}] associated with the estimated structural eigenvectors $\widehat{\bm V}_\kappa$ will be small
\begin{align*}
    \|\bm{P}_{\bm{H}} \widehat{\bm V}_\kappa\|_\F 
    &= \|\bm{P}_{\bm{H}} \widehat{\bm V}_\kappa\bm O\|_\F \\
    & = \|\bm{P}_{\bm{H}} \widehat{\bm V}_\kappa\bm O - \bm{P}_{\bm{H}} \bm V_\kappa \|_\F \\
    & = \|\bm{P}_{\bm{H}} (\widehat{\bm V}_\kappa\bm O - \bm V_\kappa)\|_\F \\
    & = \|\widehat{\bm V}_\kappa\bm O - \bm V_\kappa - \bm H \bm H^\dagger(\widehat{\bm V}_\kappa\bm O - \bm V_\kappa)\|_\F\\
    & \le \|\widehat{\bm V}_\kappa\bm O - \bm V_\kappa\|_\F + \|\bm H \bm H^\dagger\|_\F\|\widehat{\bm V}_\kappa\bm O - \bm V_\kappa\|_\F\\
    & \le (1+\sqrt{k})\frac{\sqrt{8k}\|\bm L(\estOmegaNoSup) - \bm L(\bm{\Omega})\|_2}{\Delta \lambda} \enspace ,
\end{align*}
where the first equality uses the fact that multiplication with a unitary matrix does not change the norm; 
the second equality comes from the fact that $\bm V_\kappa$ are structural eigenvectors, which means that $\bm{P}_{\bm{H}} \bm V_\kappa = \bm 0 $ because the projection error is zero;
the third equality is a simple rearrangement; 
the fourth equality follows from using the definition of the projection operator ${\bm P_{\bm H} := \bm I - \bm H \bm H^\dagger}$;
the first inequality uses the sub-additive property of the norm; 
and the final inequality follows from the Davis-Kahan theorem and the fact that $\|\bm H \bm H^\dagger\|_\F= \sqrt{k}$.
We can see from the above that $\|\bm{P}_{\bm{H}} \widehat{\bm V}\|_\F$ will be small for large enough graphs with large enough groups as $\|\bm L(\estOmegaNoSup) - \bm L(\bm \Omega)\|_2 \rightarrow 0$, which follows from the fact that the estimated entries $\widehat{{\Omega}}_{ij} \rightarrow \Omega_{ij}$ for large enough group sizes.

\section{$k$-means is the dual of minimizing projection error}
\label{sec:kmeans_proof}
Let us write out the objective function of $k$-means in which we take the rows $\bm v_{1\cdot}, \bm v_{2\cdot}, \bm \ldots,\bm v_{n\cdot}$ of $\bm V$ as $k
\times 1$ vectors representing the elements to be clustered (${\bm V^\top = [\bm v_{1\cdot}, \bm v_{2\cdot}, \bm \ldots,\bm v_{n\cdot}]}$):
\begin{equation}
    \min_{\bm H} \sum_{j=1}^k \sum_{i=1}^{n} H_{ij}\| \bm v_{i\cdot} - \bm \mu_j\|^2, \text{ with }\bm \mu_j = \frac{1}{n_j}\sum_{i=1}^n \bm v_{i\cdot} H_{ij} \enspace ,
\end{equation}
where $n_j$ is the number of points in cluster $j$.
Now observe that we can write $\bm \mu_j$ as:
\begin{equation}
    \bm \mu_j = [\bm V^\top \bm H \bm N^{-1}]_{\bm{\cdot} j} \enspace .
\end{equation}
Accordingly, we can see that $[\bm V^\top \bm H \bm N^{-1}\bm H^\top] = \bm V^\top \bm H \bm H^\dagger \in \mathbb{R}^{k\times n}$ corresponds to the matrix whose $i$-th column represents the mean of the cluster that node $i$ is assigned to. 
We can thus rewrite the $k$-means objective above as:
\begin{equation}
    \min_{\bm H} \sum_{i=1}^{n} \left \| \bm V^\top_{\bm\cdot i}- [\bm V^\top \bm H \bm H^\dagger]_{\bm\cdot i} \right \|^2.
\end{equation}
Using the Frobenius norm we can more compactly write this as:
\begin{equation}
    \min_{\bm H} \left \| \bm V^\top \!-\! \bm V^\top \bm H \bm H^\dagger \right\|_\F^2 = \min_{\bm H} \left \| \bm V^\top [\bm I - \bm H \bm H^\dagger] \right\|_\F^2.
\end{equation}
Finally, by noting that the Frobenius norm is unchanged by taking the transpose of its arguments we can establish the desired equality to our projection error criterion given in~\cref{eq:opt_projection_error}.

The above result shows that there is an striking duality between the problem of finding an EEP with minimal projection error and the $k$-means problem on the rows of the Laplacian eigenvectors: instead of searching for $k$ partition indicator vectors in an $n$-dimensional space that minimize the projection error, we can consider a dual problem of finding the $k$ centroids of $n$ points in a $k$-dimensional space (where the centroids minimize the quantization error defined via the squared 2-norm).

\section{Expected projection error}\label{app:exp_proj_error}
In this section we derive the expressions for the expected projection errors given in~\cref{eq:exp_proj_error_unconditional,eq:exp_proj_error_conditional}.

\subsection{Accounting for shared subspaces}
The derivation of $\varepsilon_0$ in Eq.~\eqref{eq:vareps0} in the main text assumes that $\bm U$ and $\bm H$ are statistically independent of each other.
However, this will not generally be true in the context of a partition indicator matrix $\bm H$ and a set of Laplacian eigenvectors $\bm V_k$.
Of particular concern is that $\bm 1$ is \emph{always} an eigenvector of a Laplacian and $\bm 1 \in \text{span}(\bm H)$ (because $\bm H \bm 1_k = \bm 1$).
We therefore need to adjust the above argument slightly to incorporate the fact that our eigenvectors will include the eigenvector $\bm 1$.
Consequently, we are actually looking for the projection of a \mbox{$(k-1)$-dimensional} (rather than $k$-dimensional) subspace in an $(n-1)-(k-1) = (n-k)$-dimensional space.
In other words, we have to account for the fact that we know that there is a one-dimensional EEP present in any connected graph. 

In general, if we know there is a $\kappa$-dimensional EEP present in the network and we are looking for the projection error associated with a set of $k>\kappa$ eigenvectors we obtain the expected error:
\begin{equation}\label{eq:exp_proj_error2}
    \varepsilon_{0}(k|\kappa) = \frac{(n-k)(k-\kappa)}{n-\kappa} \qquad \textrm{for} \quad \kappa \leq k \leq n \enspace ,
\end{equation}
which can be derived in a similar way as before by replacing $n$ with $(n-\kappa)$ and $k$ with $(k-\kappa)$ in the above derivation. 
Note that the changed denominator corresponds to the fact that the effective dimension of the space in which we calculate the projection error shrinks, since we have to exclude the shared subspace from the calculation.

To see this, let us revisit our earlier calculation and consider a random matrix of $\bm U \in \mathbb{R}^{n \times k}$ of $k$ orthonormal vectors of dimension $n$, i.e., $\bm U^\top \bm U = \bm I$.
This time, however, we will assume that we know a $\kappa$ dimensional subspace $\mathcal U_\kappa \subset \mathcal {U} = \text{im}(\bm U)$ of the space spanned by the vectors in the matrix $\bm U =[\bm u_1, \ldots, \bm u_k]$.
Without loss of generality, we may assume that we know the first $\kappa$ of these vectors and that these are simply given by standard unit vectors (otherwise we can find an orthogonal transformation $\bm Q_1$ such that $\tilde{\bm U} = \bm Q_1\bm U$ is of the desired form).
Thus we can consider a matrix $\bm U$ of the form:
\begin{equation}
    \bm U = 
    \begin{bmatrix}
        \bm I_\kappa & \bm 0_{\kappa\times(n-\kappa)}\\
        \bm 0_{(n-\kappa)\times \kappa} & \bm U_\text{sub},
    \end{bmatrix}
\end{equation}
where $\bm U_\text{sub}$ is an orthogonal matrix of size $(n-\kappa)\times(n-\kappa)$.
Following a similar calculation as above we obtain:
\begin{align}
    k &=  \E\left[\left\|\bm U\right\|_\F^2\right] = \kappa + \mathbb{E}\left[\sum_{j=1}^{k-\kappa} \sum_{i=\kappa + 1}^{n} [U_\text{sub}]_{ij}^2\right]\nonumber\\
      &= \kappa + (n-\kappa)(k-\kappa) \mathbb{E}\left[ U_{ij}^2\right]\enspace.
\end{align}
Hence, the expected value $\mathbb{E}[[\bm U_\text{sub}]_{ij}^2]$ that determines the denominator in the expected error calculation is now $1/(n-\kappa)$ instead of $1/n$.

\begin{figure}
    \centering
    \includegraphics[width=\columnwidth]{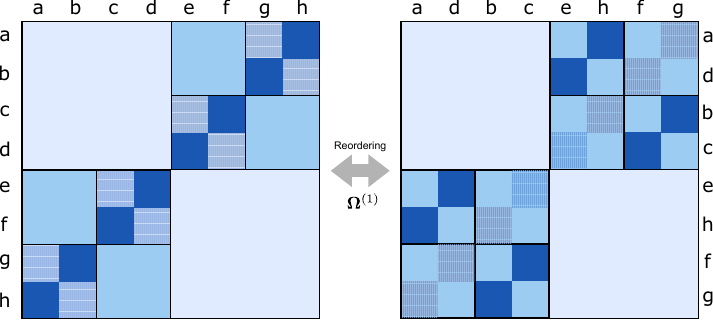}
    \caption{\textbf{Non-identifiability of hierarchical configurations for a disassortative hierarhical network model.}
    We show two (non-identifiable) possible  partitions of the affinity matrix $\bm{\Omega}^{(1)}$. The left corresponds to the ``planted'' hierarchical partition, the right to an equivalent partition, that is commensurate with a different hierarchy that preserves the coarsest and finest partition.}
    \label{fig:nonidentifiability_dis}
\end{figure}

\section{Non-identifiability for disassortative hierarchies}\label{app:identifiability}
As discussed  in \Cref{ssec:syn_results}, when trying to detect a disassortative hierarchical partition planted in a network, such the networks generated for Fig.~\ref{fig:ass_dis_exps}B \& D, we are confronted with certain non-identifiability issues. 
In such cases an algorithm can pick any of the alternative hierarchies that provide an equivalent hierarchical description, instead of the ``planted'', dissassortative hierarchy.
\Cref{fig:nonidentifiability_dis} depicts the planted hierarchical affinity matrix (\textit{left}) used in our experiments and one specific reordering of the affinity matrix (\textit{right}) that meets the nested sEEP requirement. 
Note how both the finest partition into $8$ groups as well as the coarsest partition into $2$ groups are preserved by this re-ordering. However, the two partitions into $4$ groups are inconsistent with one another. This degeneracy means that, under small perturbations of the affinity matrix, we recover a mixture of these partitions, which effectively cancels each other out and means that we often do not detect the middle level of the hierarchy.

\section{Implementation details}\label{sec:implement}
To detect the initial (finest) partition, we use the Bethe Hessian as described in~\cref{alg:BH}.

\begin{algorithm}[H]
    \SetAlgoLined
    \KwData{Adjacency matrix $\bm A$}
    \KwResult{Partition indicator matrix $\bm{H}$}
    Compute $\eta = \sqrt{\bm 1^\top \bm A \bm 1/n}$\;
    Compute $\bm B_{\pm{\eta}}$ according to~\eqref{eq:BetheHessian}\;
    Compute spectral decompositions $\bm{B}_\eta =\bm V\bm\Lambda\bm V^\top$, $\bm{B}_{-\eta} =\bm U\bm\Theta\bm U^\top$\;
    Compute $k_+ \leftarrow |\{\lambda_i : \lambda_i \le 0\}|$ and $k_{-} \leftarrow |\{\theta_i : \theta_i \le 0\}|$\;
    Estimate number of groups $\hat k = k_+ + k_{-}$\\
    Form $\bm Q = [\bm V_{k_+}, \bm U_{k_{-}}]$, containing the ($k_+$ and $k_-$) eigenvectors of $\bm{B}_{\pm \eta}$ with non-positive eigenvalues\;
    Run $k$-means clustering on the rows of $\bm Q$:
    $\bm H \leftarrow k\text{-means}(\bm Q^\top,\hat{k})$\;
    \caption{\texttt{ClusterWithBetheHessian}}%
    \label{alg:BH}
\end{algorithm}

We take the partition thus found as our finest partition and build a hierarchy by agglomeration as described in~\cref{alg:main_function}.

\begin{algorithm}[H]
    \SetAlgoLined
    \KwData{Adjacency matrix $\bm A$,\newline Initial Partition $\bm H\in \{0,1\}^{n\times k}$}
    \KwResult{Sequence of hierarchical partitions}
    $\texttt{hier\_part\_list} \leftarrow (\bm H,)$\;
    $\texttt{more\_levels} \leftarrow \texttt{true}$\;
    $u \leftarrow 1$\;
    \While{$\texttt{more\_levels}$}
    { \#Compute affinity matrix of current partition $\bm H$\;
    	$\bm \Omega \leftarrow \bm H^{\dagger(u)} \bm A \left(\bm H^{\dagger(u)} \right)^\top$; \hfill [\cref{eq:compute_omega_agglom}]\\
        \# Find (sub)partitions of $\bm \Omega$ and associated proj. errors\;
    $(\bm H_1,\ldots,\bm H_k),\bm \epsilon \leftarrow \texttt{IndentifyPartitionsAndErrors}(\bm \Omega)$\;
        \# determine candidates for hier. agglomeration\;
    $\texttt{cand\_list} \leftarrow \texttt{FindRelevantMinima}(\bm \epsilon)$\;
    \uIf{$\texttt{cand\_list} = \emptyset$}
        {\# If no agglomeration candidates exists stop
            $\texttt{more\_levels} \leftarrow \texttt{false}$\;
        }
        \Else{  \# else keep finest agg. candidate and repeat\;
            $\bm H \leftarrow \texttt{cand\_list.last}$\;
            $\texttt{hier\_part\_list.append}(\bm H)$\;
            }
    }
    \Return $\texttt{hier\_part\_list}$
    \caption{\texttt{InferHierarchy}}
    \label{alg:main_function}
\end{algorithm}

Here~\cref{alg:main_function} makes use of two subroutines. 
The first one (\cref{alg:find_part_and_errors}) creates the (best) possible subpartitions of the affinity matrix of the currently considered hierarchical level, and computes the associated projection errors.
Based on the computed projection errors we then decide whether there is evidence that there is a hierarchical refinement (\cref{alg:model_select_based_on_errors}) and keep the finest such partition. 
We then build the affinity matrix of the next hierarchical level and repeat the procedure until no more additional hierarchical levels are found.

\begin{algorithm}[H]
    \SetAlgoLined
    \KwData{Affinity matrix $\bm \Omega \in \mathbb{R}^{k\times k}$,\newline number of perturbed samples $z$}
    \KwResult{Sequence of partitions $(\bm H_1,\ldots,\bm H_k)$, \\ \qquad \quad mean projection errors $\bm \epsilon$}
    \# Compute Laplacian\\
    $\bm L \leftarrow \text{diag}(\bm \Omega \bm 1) - \bm \Omega$\;
    \# Create uniform random walk matrix $\bm W$\\ 
    $\bm W \leftarrow \bm I - \frac{1}{d_\text{max}} \bm L$ \hfill [\cref{eq:uniform_rw}]\\
    Compute spectral decomposition of $\bm W$:\newline
    $\bm  W \leftarrow \bm V \bm \Lambda \bm V^\top$ with $|\lambda_1| = 1 >\ldots>|\lambda_k|$\;
    \For{$r = 2:k-1$}
    {\# Assemble matrix of first $r$ eigenvectors\\ 
    	$\bm V_r \leftarrow \bm V_{\bm\cdot,[1:r]}$ \;
        \# Find partition into $r$ groups using $k$-means\\ $\bm H_{r} \leftarrow \text{k-means}(\tilde{\bm V}_r^\top)$\;
    }
    \# for $r=1$ and $r=k$ there is only one possibe partition\\
    $\bm  H_1 \leftarrow \bm 1$, $\bm H_k \leftarrow \bm I_k$\;
    \For{$\zeta = 1:z$}
    {$\tilde{\bm{\Omega}} \leftarrow \text{perturbation}(\bm\Omega)$\;
    \# Compute Laplacian\;
    $\tilde{\bm L} \leftarrow \text{diag}(\tilde{\bm \Omega} \bm 1) - \tilde{\bm \Omega}$\;
    \# Create uniform random walk matrix $\widetilde{\bm W}$\; 
    $\widetilde{\bm W} \leftarrow \bm I - \frac{1}{d_\text{max}} \tilde{\bm L}$ \hfill [\cref{eq:uniform_rw}]\\
    \# Compute spectral decomposition of $\widetilde{\bm W}$:\\
    $\widetilde{\bm  W} \leftarrow \bm U \bm \Theta \bm U^\top$ with $|\theta_1| = 1 >\ldots>|\theta_k|$\;
    \For{$r = 1:k$}
        {
            \# Assemble matrix of first $r$ eigenvectors:\\ 
            $\bm U_r \leftarrow \bm U_{\bm\cdot,[1:r]}$\;
        \# Compute projection error\\ 
		$\bm P_{\bm{H}_r} \leftarrow [\bm I -\bm H_r^{}\bm H_r^\dagger]$; \hfill [\cref{eq:projection_onto_H}]\\
    	$\varepsilon^{(\zeta)}(r) \leftarrow  \|\bm{P}_{\bm{H_r}} \bm U_r \|_\F^2$; \hfill [\cref{eq:proj_error}]\\        
        }
    }
    \# Compute mean projection error vector (over $z$ samples)\\
    $\epsilon_r \leftarrow  \frac{1}{z}\sum_{\zeta} \bm \varepsilon^{(\zeta)}(r)$\;
    \Return $(\bm H_1,\ldots,\bm H_k),\bm \epsilon$
\caption{\texttt{IdentifyPartitionsAndErrors}}\label{alg:find_part_and_errors}
\end{algorithm}

\pagebreak

\begin{algorithm}
    \SetAlgoLined
    \KwData{Mean projection error $\bm \epsilon$}
    \KwResult{Indices of hierarchical partition candidates}
    \# get size of projection error vector\\ \# equal to maximal number of groups\\
    $k = \text{length}(\bm \epsilon)$\;
    $\texttt{cand\_list} \leftarrow \emptyset$\;
	\# Compute expected error $\bm \varepsilon_0$\\
    $\varepsilon_0(r) \leftarrow \frac{(n-r)(r-1)}{n-1}$; \hfill [\cref{eq:exp_proj_error_unconditional}]\\
    \# Calculate Mean squared logistic error (MSLE)\\
    $ \text{MSLE}_0 \leftarrow \min_{\sigma} \frac{1}{k} \sum_{r=1}^k \left(
  \log \frac{\left[\epsilon_r + 1\right]}{\left[\sigma \varepsilon_0(r) + 1 \right]} 
  \right)^2$ \hfill [\cref{eq:MSLE}]\\ 
  \For{$\kappa = 2:k-1$}{
      \# Get candidate levels and compute expected error\\
    $\kappa \leftarrow \text{cand\_list}$\;
	Calculate $\varepsilon_{0}(r|\kappa)$ according to \hfill [\cref{eq:exp_proj_error_cond_general}]\\
	\# Calculate Mean squared logistic error (MSLE)\\
    $ \text{MSLE}_{\kappa} \leftarrow \min_{\sigma} \frac{1}{k} \sum_{r=1}^k \left(
  \log \frac{\left[\epsilon_r + 1\right]}{\left[\sigma \varepsilon_0(r|\kappa) + 1 \right]} 
  \right)^2$ \hfill [\cref{eq:MSLE}]\\   
  \If{$\textrm{MSLE}_{\kappa} < \text{MSLE}_{0}$}
  {$\texttt{cand\_list.append}(\kappa)$\;
      $\text{MSLE}_0 \leftarrow \text{MSLE}_\kappa$\; }
     }
    \Return $\texttt{cand\_list}$
    \caption{$\texttt{FindRelevantMinima}$}%
    \label{alg:model_select_based_on_errors}
\end{algorithm}

\begin{algorithm}
    \SetAlgoLined
    \KwData{Adjacency Matrix $\bm A$}
    \KwResult{Partition indicator matrix $\bm H$}
    Compute $k_+, k_{-}$ according to Alg.~\ref{alg:BH}\;
    Estimate number of groups $\hat k = k_+ + k_{-}$\;
    Estimate the spectral radius $ \rho = \frac{\bm 1^\top \bm A \bm A \bm 1/n}{\bm 1^\top \bm A \bm 1/n} $\;
    \For{$i = 2:k_+$}{
    	\# Find $\eta \in (1, \sqrt{\rho})$ such that $\lambda_i(B_{+\eta})=0$ \\
        $\zeta_i \leftarrow \eta$\;
        \# Compute spectral decomposition $\bm{B}_{\zeta_i} =\bm V\bm\Lambda\bm V^\top$\;
        $\bm V^+_i \leftarrow \bm v_i(B_{+\zeta_i})$\;
    }
    \For{$i = 1:k_-$}{
        \# Find $\eta \in (1, \sqrt{\rho})$ such that $\lambda_i(B_{-\eta})=0$ \\
        $\zeta_i \leftarrow \eta$\;
        \# Compute spectral decomposition $\bm{B}_{-\zeta_i} =\bm V\bm\Lambda\bm V^\top$\;
        $\bm V^-_i \leftarrow \bm v_i(B_{-\zeta_i})$\;
    }
    Form $\bm V^+=[\bm V^+_2, \bm V^+_3, ..., \bm V^+_{k_+}]$, $\bm V^-=[\bm V^-_1, \bm V^-_2, ..., \bm V^-_{k_-}]$\;
    Form $\bm Q=[\bm V^+, \bm V^-]$\;
    \# Run k-means clustering on rows of $\bm Q$:\\
    \textbf{H} $\leftarrow$ \text{k-means}($\bm Q^T$, $\hat k-1$)\;
    \caption{$\texttt{ClusterWithDegreeCorrectedBetheHessian}$}
    \label{alg:DC_BH}
\end{algorithm}

\end{document}